\documentclass[preprint2]{aastex}
\newcommand{\EQ}{\begin{equation}}
\newcommand{\EN}{\end{equation}}
\newcommand{\EQA}{\begin{eqnarray}}
\newcommand{\ENA}{\end{eqnarray}}
\newcommand{\eq}[1]{(\ref{#1})}

\newcommand{\Eq}[1]{Eq.~(\ref{#1})}
\newcommand{\Eqs}[2]{Eqs~(\ref{#1}) and~(\ref{#2})}

\newcommand{\eqs}[2]{(\ref{#1}) and~(\ref{#2})}

\newcommand{\Sec}[1]{\S\ref{#1}}
\newcommand{\Secs}[2]{\S\S\ref{#1} and \ref{#2}}
\newcommand{\Fig}[1]{Fig.~\ref{#1}}

\newcommand{\Tab}[1]{Table~\ref{#1}}

\newcommand{\bra}[1]{\langle #1\rangle}

\newcommand{\meanemf}{\overline{\mbox{\boldmath ${\cal E}$}} {}}
\newcommand{\meanAA}{\overline{\bf{A}}}
\newcommand{\meanBB}{\overline{\bf{B}}}
\newcommand{\meanJJ}{\overline{\bf{J}}}
\newcommand{\meanUU}{\overline{\bf{U}}}
{}
{}
{}
{}
{}
{}
{}
\newcommand{\qq}{\mbox{\boldmath $q$} {}}

\newcommand{\uu}{{\bf{u}}}
\newcommand{\BB}{{\bf{B}}}
\newcommand{\JJ}{{\bf{J}}}
\newcommand{\jj}{{\bf{j}}}
\newcommand{\AAA}{{\bf{A}}}
\newcommand{\aaaa}{{\bf{a}}}
\newcommand{\bb}{{\bf{b}}}

\newcommand{\nab}{\mbox{\boldmath $\nabla$} {}}

\newcommand{\oo}{\mbox{\boldmath $\omega$} {}}

\newcommand{\emf}{\mbox{\boldmath ${\cal E}$} {}}

\newcommand{\ii}{{\rm i}}

\newcommand{\DD}{{\rm D} {}}
\newcommand{\dd}{{\rm d} {}}

\def\ga{\mathrel{\mathchoice {\vcenter{\offinterlineskip\halign{\hfil
$\displaystyle##$\hfil\cr>\cr\sim\cr}}}
{\vcenter{\offinterlineskip\halign{\hfil$\textstyle##$\hfil\cr>\cr\sim\cr}}}
{\vcenter{\offinterlineskip\halign{\hfil$\scriptstyle##$\hfil\cr>\cr\sim\cr}}}
{\vcenter{\offinterlineskip\halign{\hfil$\scriptscriptstyle##$\hfil\cr>\cr\sim\cr}}}}}
\newcommand{\ea}{{\rm et al.\ }}

\def\half{{\textstyle{1\over2}}}

\def\onethird{{\textstyle{1\over3}}}

\newcommand{\yapj}[3]{ #1, {ApJ,} {#2}, #3}
\newcommand{\yapjl}[3]{ #1, {ApJ,} {#2}, #3}

\newcommand{\yan}[3]{ #1, {AN,} {#2}, #3}

\newcommand{\yana}[3]{ #1, {A\&A,} {#2}, #3}

\newcommand{\yjfm}[3]{ #1, {JFM,} {#2}, #3}
\newcommand{\ypf}[3]{ #1, {Phys. Fluids,} {#2}, #3}
\newcommand{\ypp}[3]{ #1, {Phys. Plasmas,} {#2}, #3}

\newcommand{\yprl}[3]{ #1, {PRL,} {#2}, #3}

\newcommand{\ymn}[3]{ #1, {MNRAS,} {#2}, #3}

\newcommand{\ypr}[3]{ #1, {Phys. Rev.,} {#2}, #3}

\newcommand{\yjour}[4]{ #1, {#2}, {#3}, #4}

\newcommand{\ybook}[3]{ #1, {#2} (#3)}
\newcommand{\yproc}[5]{ #1, in {#3}, ed. #4 (#5), #2}

\newcommand{\pjour}[2]{ #1, {#2,} (in press)}

\begin{document}

\title{Dynamic nonlinearity in large scale dynamos with shear}
\author{Eric G.\ Blackman\altaffilmark{1} and Axel Brandenburg\altaffilmark{2}}
\affil{$^1$ Department of Physics \& Astronomy, University of Rochester, Rochester NY 14627}
\affil{$^2$ NORDITA, Blegdamsvej 17, DK-2100 Copenhagen \O, Denmark}
\date{\today,~ $ $Revision: 1.146 $ $}

\begin{abstract}
We supplement the mean field dynamo growth equation 
with the total magnetic helicity evolution equation. 
This provides an explicitly time dependent model for
alpha quenching in dynamo theory.
For dynamos without shear, this approach
accounts for the observed large scale field growth and
saturation in numerical simulations. After a significant kinematic phase, the 
dynamo is resistively quenched, i.e.\ the saturation time depends on the
microscopic resistivity. This is independent of whether or not the turbulent 
diffusivity is resistively quenched.
We find that the approach is also successful for dynamos
that include shear and exhibit migratory waves (cycles). 
In this case however, 
whether or not the cycle period remains of the order of the 
dynamical time scale at large magnetic Reynolds numbers does depend how 
on how the turbulent magnetic diffusivity quenches. Since this
is unconstrained by magnetic helicity conservation, the 
diffusivity is  presently an input parameter.   Comparison to  
current numerical experiments  suggests a turbulent diffusivity that
depends only weakly on the magnetic Reynolds number, but 
higher resolution simulations are needed.
\end{abstract}


\section{Introduction}

The large scale magnetic field of the sun and other stars is frequently
modeled using $\alpha\Omega$ dynamo theory (Moffatt 1978; Parker 1979;
Krause, \& R\"adler 1980; Zeldovich, Ruzmaikin, \& Sokoloff 1983).
This theory has been successful in reproducing the cyclic behavior
of solar and stellar activity as well as the latitudinal migration of
belts of magnetic activity. The cyclic behavior results mainly from the
shear (the $\Omega$ effect); see the references above. Shear also helps
producing a strong toroidal field, while the $\alpha$ effect remains
responsible for regenerating poloidal from toroidal field. Dynamos
without (or weak) shear can also generate large scale fields, but now the
$\alpha$ effect also regenerates toroidal from poloidal field ($\alpha^2$
dynamo). Such dynamos are usually nonoscillatory. On the other hand,
not all $\alpha\Omega$ dynamos are oscillatory: in oblate geometries
(accretion discs and galaxies), dynamos tend to be nonoscillatory
(e.g., Covas et al.\ 1999). In simple cartesian geometry with periodic
boundaries $\alpha$ effect dynamos always exhibit migratory waves once
shear is strong enough.

The viability of an $\alpha$ effect dynamo has been controversial 
primarily because the nonlinear backreaction 
of the growing magnetic field on the dynamo coefficients
has not been well understood.
At the center of the debate is how to incorporate the backreaction into
the $\alpha$ effect. It is
often taken to be of the form $\alpha=\alpha_{\rm K}\,q(\meanBB)$,
where $\alpha_{\rm K}$ is the kinematic value and
\EQ
q=\left(1+a\meanBB^2/B_{\rm eq}^2\right)^{-1}
\label{lorentzian}
\EN
is a lorentzian quenching function.
Here, $\meanBB^2$ is the mean field,
$B_{\rm eq}$ is the equipartition field strength and $a$ is a
dimensionless parameter. In recent years there has been mounting
concern that the value of $a$ might actually be of the order of
the magnetic Reynolds number, $R_{\rm m}$ (Vainshtein \& Cattaneo 1992;
Gruzinov \& Diamond 1994; 1995; 1996; Bhattacharjee \& Yuan 1995;
Cattaneo \& Hughes 1996), rather than of order unity
(e.g., R\"udiger, \& Kitchatinov 1993). In stars, $R_{\rm m}\sim10^{8\ldots9}$,
so quenching would set in for rather weak fields, suggesting
that dynamo-generated fields should be much below the equipartition field
strength. This is therefore referred to as `catastrophic' quenching.

However, quenchings of the form \eq{lorentzian} have traditionally been 
obtained under the assumption that the system is in a steady state.
This leads to incorrect predictions about the evolution
of the mean field. Since the magnetic helicity evolution equation 
must be taken into account, and because it is time dependent,
the functional form for $\alpha$ quenching also becomes time dependent.
Loosely speaking, a helical field is one with a field-aligned current.
While this characterizes primarily the current helicity density, the
magnetic helicity is really a volume integral which can be associated
with the topological linkage of flux lines. Both magnetic helicity and
topological linkage are conserved in the non-resistive (large magnetic
Reynolds number) limit. This imposes a crucial constraint on the
field evolution in general, and the evolution of the $\alpha$ effect
in particular.

There have been a number of important attempts to incorporate dynamical  
$\alpha$ quenching based on magnetic helicity conservation
(Kleeorin \& Ruzmaikin 1982; Zeldovich et al.\ 1983;
Kleeorin, Rogachevskii, \& Ruzmaikin 1995; see also Seehafer 1996; Ji 1999).
Recently  Field \& Blackman (2002, hereafter FB02)
derived from Pouquet, Frisch and L\'eorat (1976), 
a simple two-scale approach 
whose solutions and physical interpretation were shown 
to agree well with simulations of Brandenburg (2001a, hereafter B01).

Let us define some terms:  
we refer to the procedure of obtaining $\alpha$ quenching from
a time dependent differential equation derived from magnetic helicity
conservation as ``dynamical'' quenching. 
In this case, $\alpha$ is obtained by solving an explicitly time dependent
equation. On the other hand, we refer to the quenching as fixed form
or ``algebraic'' if
$\alpha$ is expressed as a fixed function 
of $\BB$, where by fixed we mean that 
the functional form is kept fixed in time,
and the only time dependence enters through $\BB$. 
Although certain algebraic quenching expressions can emerge as a useful
approximation in specific temporal regimes, 
we shall show below that only
the dynamical quenching is consistent with the magnetic
helicity equation, and the functional form of $\alpha(\BB)$ must
change with time. The lorentzian quenching formula (\ref{lorentzian}), when
applied for all time,  would be an example of a fixed form 
algebraic quenching prescription.

B01 performed three-dimensional simulations of a zero shear helical 
dynamo  in a periodic domain and found that the field attains
super-equipartition field strengths.
The initial growth phase is kinematic, and it
is only at later times that a saturation phase emerges.
The final super-equipartition field strength
is reached after a resistive time scale.
This behavior can be
reproduced empirically by an $\alpha^2$ dynamo where both $\alpha$ and
turbulent magnetic diffusivity, $\eta_{\rm t}$, are catastrophically
quenched according to \Eq{lorentzian} with $a\sim R_{\rm m}$ (B01).
Below we show however that this form of the turbulent electromotive
force is not universal. The dynamical expression is degenerate in
special cases where the mean field is current-free or force-free;
it therefore reproduces catastrophically quenched behavior in
those cases, but not in others.

For an $\alpha^2$ dynamo in a periodic box,
growth of the large scale magnetic field energy 
is associated directly with the growth of large scale magnetic helicity.
Since the total magnetic helicity can only change resistively,
growth of the large scale magnetic helicity implies 
significant growth of small scale magnetic helicity of the opposite sign.
By taking $\alpha$ as proportional
to the difference between kinetic and current helicities
[the ``relative helicity'' as derived 
by Pouquet et al.\ (1976)] and using a two-scale approach,
FB02 developed a model for the nonlinear
dynamical quenching of $\alpha$, whose equation for $\alpha$ 
is formally equivalent to that of Kleeorin \& Ruzmaikin (1982)
and Zeldovich et al.\ (1983). 
Using a two-scale approach, FB02 showed    
that the  growth of the small scale magnetic helicity 
augments the current helicity contribution to $\alpha$ which ultimately
quenches it. The coupling between the small and large scale magnetic
helicity equations in this 
two-scale time dependent dynamical quenching theory
predicts that at late times, the quenching of $\alpha$ is consistent
with \Eq{lorentzian} with $a\sim R_{\rm m}$, 
but that at early times the dynamo proceeds kinematically, independent
of $R_{\rm m}$. The results agree
with the simulations of B01 both in terms of the time
evolution of the large scale field energy and the saturation value.

An explicitly time dependent quenching formula
has sometimes been used in mean-field models,
but the main motivation in many instances was to 
study chaotic behavior in
stellar dynamos (Ruzmaikin 1981; Schmalz \& Stix 1991;
Feudel, Jansen, \& Kurths 1993; Covas \ea 1997; 1998).
Some kind of dynamical $\alpha$ quenching, but with an explicit time lag,
has previously been invoked by Yoshimura (1978) in order to reproduce long
term behavior with multiple periods. This approach was however rather ad
hoc and not based on magnetic helicity conservation.
It was only recently that Kleeorin \ea (1995; 2000) pointed out that
the catastrophic quenching of Vainshtein \& Cattaneo (1992)
can emerge from the dynamical quenching approach under
certain circumstances (e.g.\ when the mean field is current-free
or force-free).

The plan of the paper is as follows. In \Secs{Sformalism}{Scomplete}
we present the governing equations in a form most suitable for
our analytic and numerical treatment. We then consider limiting cases
such as early and late time evolution, the effective $\alpha$ during
the different stages, and the effects of shear (\Sec{SPrelim}).
We then consider the full time evolution numerically with and without
shear, and compare with fully three-dimensional simulations
(\Sec{Sfull}). Unlike dynamos without shear, dynamos with shear can
exhibit dynamo waves and thus cyclic behavior. 
In addition to being a distinction of astrophysical
relevance, we will see that this distinction 
is important in assessing the role of turbulent diffusion.
Finally, we present possible extensions of the model
(\Sec{Sextent}) and present our conclusions (\Sec{Sconcl}).

\section{The dynamical equation for $\alpha$}
\label{Sformalism}

We use the magnetic helicity equation for the fluctuating field as
an auxiliary equation that needs to be solved simultaneously with
the mean-field dynamo equation. For the $\alpha^2$ dynamo case, 
the dynamical quenching theory derived below is similar to
that of FB02, but here we generalize the approach to the
$\alpha\Omega$ dynamo.

In a closed or periodic domain the magnetic helicity,
$\bra{\AAA\cdot\BB}$, evolves according to
\EQ
{\dd\over\dd t}\bra{\AAA\cdot\BB}=-2\eta\mu_0\bra{\JJ\cdot\BB},
\label{helicity_eqn}
\EN
where $\AAA$ (with $\BB=\nab\times\AAA$) is the magnetic vector potential,
$\JJ=\nab\times\BB/\mu_0$ is the current density, $\eta$ is the microscopic
magnetic diffusivity, and angular brackets denote volume averages.
We split the magnetic field into mean and fluctuating components, i.e.\
$\BB=\meanBB+\bb$ (and similarly for all other quantities). Mean fields
are here defined by averaging over one or two coordinate directions,
depending on whether the mean field is two- or one-dimensional; see below.
The evolution
of the mean magnetic vector potential is given by
\EQ
{\partial\meanAA\over\partial t}=\meanemf+\meanUU\times\meanBB
-\eta\mu_0\meanJJ-\nab\overline{\phi},
\label{dAdt}
\EN
where $\meanemf=\overline{\uu\times\bb}$ is the electromotive force resulting
from small scale velocity and magnetic fields, and $\overline{\phi}$ is the
electrostatic potential of the mean field which can be chosen arbitrarily without
affecting the magnetic field and magnetic helicity. From \Eq{dAdt} one obtains
an evolution equation for the magnetic helicity of the mean field,
\EQ
{\dd\over\dd t}\bra{\meanAA\cdot\meanBB}=
2\bra{\meanemf\cdot\meanBB}-2\eta\mu_0\bra{\meanJJ\cdot\meanBB},
\label{dABdt}
\EN
and an evolution equation for the magnetic helicity of the fluctuating
field,
\EQ
{\dd\over\dd t}\bra{\aaaa\cdot\bb}=
-2\bra{\meanemf\cdot\meanBB}-2\eta\mu_0\bra{\jj\cdot\bb},
\label{dabdt}
\EN
such that the sum of the two equations becomes \Eq{helicity_eqn}. 
We note that $\meanUU$ does not enter \Eqs{dABdt}{dabdt}.
A remarkable property of \Eq{dabdt} is that it contains no triple moments,
in contrast to the energy equation for the fluctuating magnetic field,
for example.
This property allows a closure whereby \Eq{dabdt} is solved along with
the mean-field equations to ensure that the magnetic helicity equation
\eq{helicity_eqn} is satisfied exactly.

We now discuss the functional form of $\meanemf$.
In mean-field electrodynamics one can show that for isotropic
homogeneous turbulence
(Moffatt 1978)
\EQ
\meanemf
=\alpha\meanBB-\eta_{\rm t}\mu_0\meanJJ,
\label{emf}
\EN
where $\alpha$ will be specified below and
$\eta_{\rm t}$ is the turbulent diffusivity.
The anisotropies induced by the generated large scale field are ignored.
In \Eq{emf} we have
ignored a possible contribution from the cross helicity effect
(Yoshizawa \& Yokoi 1993); in what follows, we assume that
the small scale cross helicity is always small.
We have measured its contribution to $\emf$ in a simulation with shear and
found it to be $\sim1/20$ of the contribution from the $\alpha$ effect.

To proceed, we take the nonlinear $\alpha$ of the form 
originally proposed in Pouquet et al.\ (1976),
where the {\it residual} (sum of kinetic and
magnetic) isotropic and homogeneous $\alpha$ effect,
\EQ
\alpha=\alpha_{\rm K}+\alpha_{\rm M}.
\label{alpha}
\EN
is given by
\EQ
\alpha_{\rm K}=-\onethird\tau\bra{\oo\cdot\uu},\quad
\alpha_{\rm M}=+\onethird\tau\bra{\jj\cdot\bb}/\rho_0.
\label{alphaKM}
\EN
Angular brackets denote volume averages. (This implies that these and
other turbulent transport coefficients are constant in space.)
Non-isotropic tensorial generalizations to \eq{alphaKM} are known
(Kleeorin \& Rogachevskii 1999; Rogachevskii \& Kleeorin 2001).
In \Eq{alphaKM}, $\tau$ is the
correlation time, $\bra{\oo\cdot\uu}$ is the small scale kinetic helicity (with
$\oo=\nab\times\uu$), $\bra{\jj\cdot\bb}$ is the small scale current helicity
(with $\jj=\nab\times\bb/\mu_0$, where $\mu_0$ is the vacuum
permeability), and $\rho_0$ is the density which is assumed constant.

To account for the magnetic influence
on the turbulent magnetic diffusivity, we assume the form
\EQ
\eta_{\rm t}=\eta_{\rm t0}g(\meanBB),
\label{eta}
\EN
where
\EQ
\eta_{\rm t0}=\onethird\tau\bra{\uu^2}
\label{etaK}
\EN
is the kinematic value
of the turbulent magnetic diffusivity, and $g(\meanBB)$ is a
quenching function (specified later) normalized such that $g(0)=1$.
We use \Eq{etaK} to eliminate $\tau$ in \Eq{alphaKM}, so
\EQ
\alpha_{\rm M}=\eta_{\rm t0}\mu_0{\bra{\jj\cdot\bb}/B_{\rm eq}^2}.
\label{alphaKM2}
\EN
with $B_{\rm eq}^2=\mu_0\rho_0\bra{\uu^2}$.\footnote{In principle,
the effective correlation times in the expressions for $\alpha_{\rm M}$
and $\eta_{\rm t0}$ ($\tau_{\rm M}$ and $\tau_{\rm K}$, say) could be
different. This would correspond to replacing
$B_{\rm eq}^2\rightarrow(\tau_{\rm M}/\tau_{\rm K})B_{\rm eq}^2$
in the final expressions involving $B_{\rm eq}$.}

In isotropic turbulence, the spectra of magnetic and current helicities
are related to each other by a $k^2$ factor where $k$ is the wavenumber.
To a good approximation
this also applies in real space to the helicities at the two
scales of the fluctuating and mean fields. In particular, we have
\EQ
\bra{\aaaa\cdot\bb}=\mu_0\bra{\jj\cdot\bb}/k_{\rm f}^2
=\alpha_{\rm M}{B_{\rm eq}^2/(\eta_{\rm t0}k_{\rm f}^2)},
\label{ab_to_alpM}
\EN
where we have used \Eq{alphaKM2} to relate
$\bra{\aaaa\cdot\bb}$ to $\alpha_{\rm M}$. Here, $k_{\rm f}$
is the characteristic wavenumber of the fluctuating field.
After multiplying
\Eq{dabdt} by $\eta_{\rm t0}k_{\rm f}^2/B_{\rm eq}^2$ we obtain
an evolution equation for $\alpha_{\rm M}$ (Kleeorin \& Ruzmaikin 1982; see also
Zeldovich et al.\ 1983; and Kleeorin et al.\ 1995)
\EQ
{\dd\alpha_{\rm M}\over\dd t}=-2\eta_{\rm t0}k_{\rm f}^2\left(
{\bra{\meanemf\cdot\meanBB}\over B_{\rm eq}^2}
+{\alpha_{\rm M}\over R_{\rm m}}\right),
\label{dynquench}
\EN
where we have defined the magnetic Reynolds number as
\EQ
R_{\rm m}=\eta_{\rm t0}/\eta.
\EN
This result agrees with that in Kleeorin et al.\ (1995) if their
characteristic length scale of the turbulent motions at the
surface, $l_{\rm s}$, is identified with
$2\pi/k_{\rm f}$ and if their parameter $\mu$ is identified
with $8\pi^2\eta_{\rm t0}^2/(\bra{\uu^2}l_{\rm s}^2)$.
Note that  solving \Eq{dAdt} or \eq{dABdt} together with
\eq{dynquench} 
is equivalent to solving \Eq{dAdt} or \eq{dABdt} with
\eq{dabdt}; FB02 solved \eq{dABdt} with \eq{dabdt}
for the $\alpha^2$ dynamo.

The quenching function for magnetic diffusivity, $g(\meanBB)$, is
uncertain. Cattaneo \& Vainshtein (1991) proposed a catastrophic quenching
formula for $g(\meanBB)$ in two-dimensional turbulence. Gruzinov
\& Diamond (1994) confirmed this, but found no quenching in the
three-dimensional case, i.e.\ $g=1$ for all field strengths. This is in
qualitative agreement with numerical simulations of Nordlund, Galsgaard,
\& Stein (1994). Kitchatinov, R\"udiger, \& Pipin (1994)
as well as Rogachevskii \& Kleeorin (2001) found that
$g\propto|\meanBB|^{-1}$ for strong fields and independent of
$R_{\rm m}$. Instead of using their detailed functional (and
tensorial) formulations, we adopt here a simple fit formula 
\EQ
g=(1+\tilde{g}|\bra{\meanBB}|/B_{\rm eq})^{-1}\quad\mbox{(case I)},
\label{g}
\EN
which was also used in B01 who found $\tilde{g}\approx16$ for runs
with different values of $R_{\rm m}$. This formula matches the
asymptotic form of Eq.~(20) of Rogachevskii \& Kleeorin (2001)
with $\tilde{g}=5\sqrt{2}/\pi\approx2.25$ if
$\bra{\bb^2}\approx\mu_0\rho_0\bra{\uu^2}$ is assumed
(but it varies only little with the level of small scale magnetic
energy: $\tilde{g}=2.78$ if $\bra{\bb^2}=0$ is assumed, for example).
In the following we allow for different values of $\tilde{g}$,
including $\tilde{g}=0$. We emphasize that our prescription for 
the quenching of
$\eta_{\rm t}$ is not dynamical, because the quenching depends
on $\bra{\bb^2}$ which does not obey such
a stringent conservation law as $\bra{\aaaa\cdot\bb}$,
as does that governing $\alpha$.
Nevertheless, for fully helical fields, $\bra{\bb^2}$ and
$\bra{\aaaa\cdot\bb}$ are proportional to each other. This, as well
as earlier work by B01 and FB02 motivates use of the expression
\EQ
g=\alpha/\alpha_{\rm K}\quad\mbox{(case II)},
\EN
which will also be considered below for comparison.

\section{The complete set of model equations}
\label{Scomplete}

To summarize our approach, the problem consists of simultaneously solving
the two equations
\EQ
{\partial\meanBB\over\partial t}=\nab\times
\left(\meanUU\times\meanBB+\alpha\meanBB
-(\eta+\eta_{\rm t})\mu_0\meanJJ\right),
\label{fullset1}
\EN
\EQ
{\dd\alpha\over\dd t}=-2\eta_{\rm t0}k_{\rm f}^2\left(
{\alpha\bra{\meanBB^2}-\eta_{\rm t}\mu_0\bra{\meanJJ\cdot\meanBB}
\over B_{\rm eq}^2}+{\alpha-\alpha_{\rm K}\over R_{\rm m}}\right),
\label{fullset2}
\EN
where $\eta_{\rm t}$ depends on $\meanBB$ via
\Eq{eta}. The $\eta_{\rm t0}$ coefficient in \Eq{fullset2}
is however constant. (In practice we continue solving for $\meanAA$
instead of $\meanBB$.) We emphasize
that $\meanBB$ is spatially periodic and $\alpha$ and $\eta_{\rm t}$
are spatially uniform. The generalization to non-periodic $\meanBB$
and spatially varying forms of $\alpha$ and $\eta_{\rm t}$ is not
straightforward and will be discussed at the end of the paper.

Shear (which models differential rotation) can be implemented in the
form $\meanUU=(0,Sk_1^{-1}\cos k_1 x,0)$, where $k_1$ is the
minimum wavenumber. In this case the mean field
is two-dimensional, i.e.\ $\meanBB=\meanBB(x,z,t)$, and comparison
with corresponding turbulence simulations of 
Brandenburg, Bigazzi, \& Subramanian (2001, hereafter BBS)
Brandenburg, Dobler, \& Subramanian (2002, hereafter BDS) is
possible. A simpler model, that we will also consider, is one with
linear shear, $\meanUU=(0,Sx,0)$. In that case the mean field is
one-dimensional, i.e.\ $\meanBB=\meanBB(z,t)$. Corresponding
turbulence simulations can be carried out in the shearing
sheet approximation, which allows pseudo-periodic boundary
conditions in $x$. In that case, however, there are currently only the
more complicated simulations relevant to accretion discs
(Brandenburg et al.\ 1995), so comparison with the present work
is difficult.

The dynamo efficiency is determined by the usual dynamo parameters
\EQ
C_\alpha=\alpha_{\rm K}/(\eta_{\rm T0}k_1),\quad
C_\Omega=S/(\eta_{\rm T0}k_1^2),
\EN
where $\eta_{\rm T0}=\eta+\eta_{\rm t0}$.
In addition we have to specify $R_{\rm m}$, $k_{\rm f}$, and a parameter
$\epsilon_{\rm f}$ which measures the degree to which the small scale
field is helical (defined in the next section). As a non-dimensional
measure of $k_{\rm f}$ we introduce $\kappa_{\rm f}=k_{\rm f}/k_1$.
Furthermore, in the absence of a
satisfactory theory for $\eta$ quenching, the parameter $\tilde{g}$
also has to be specified. The problem is therefore completely
described by the 6 dimensionless parameters $C_\alpha$, $C_\Omega$,
$R_{\rm m}$, $k_{\rm f}$, $\epsilon_{\rm f}$,
and $\tilde{g}$.

When comparing with simulations we may use the
rough estimate $R_{\rm m}\approx u_{\rm rms}/(\eta k_{\rm f})$.
A crude estimate for $C_{\alpha}$ can be obtained using \Eqs{alphaKM}{etaK}
together with $\bra{\oo\cdot\uu}\approx\tilde{k}_{\rm f}\bra{\uu^2}$,
where $\tilde{k}_{\rm f}=\epsilon_{\rm f}k_{\rm f}$,
so $\alpha_{\rm K}\approx\tilde{k}_{\rm f}\eta_{\rm t0}$.
It is then convenient to define the nondimensional effective wavenumber of
the fluctuating field, $\tilde{\kappa}_{\rm f}=\tilde{k}_{\rm f}/k_1$, so
\EQ
C_{\alpha}\approx\tilde{\kappa}_{\rm f}/\iota,
\label{Calp}
\EN
where we have introduced the correction factor
\EQ
\iota\equiv\eta_{\rm T0}/\eta_{\rm t0}=1+R_{\rm m}^{-1},
\EN
where $\eta_{\rm T0}=\eta_{t0}+\eta$.
This ratio is just above unity for $R_{\rm m}\gg1$.

In addition to one- and two-dimensional models,
we shall also consider a one-mode reduction that reduces
the one-dimensional vector equation for $\meanAA$ (or equivalently
for $\meanBB$) to 2 ordinary (complex) differential equations.  
This procedure is similar, although more general, than that of FB02 where 
only two ordinary differential 
equations \Eq{dABdt} and \eq{dynquench} were solved
for the $\alpha^2$ maximally helical dynamo.  In our present case,
non-maximally helical dynamos with shear can also be studied.

\section{Preliminary considerations}
\label{SPrelim}

\subsection{Final field strength}
\label{Sfinal}

For helical (or partially helical) fields, the resulting
steady-state field strength is determined by $\bra{\JJ\cdot\BB}=0$;
see \Eq{helicity_eqn}. In terms of mean and fluctuating fields this means
\EQ
\bra{\meanJJ\cdot\meanBB}=-\bra{\jj\cdot\bb};
\label{Eq41ofB01}
\EN
see Eq.~(41) of B01.

In order to connect the current helicities with magnetic energies,
we can now define the
effective wavenumbers for mean and fluctuating fields
more precise and write,
\EQ
\tilde{k}_{\rm m}=k_{\rm m}\epsilon_{\rm m}
=\mu_0\bra{\meanJJ\cdot\meanBB}/\bra{\meanBB^2},
\label{kmtilde}
\EN
\EQ
\tilde{k}_{\rm f}=k_{\rm f}\epsilon_{\rm f}=\mu_0\bra{\jj\cdot\bb}/\bra{\bb^2}.
\label{kftilde}
\EN
Here, $k_{\rm m}$ and $k_{\rm f}$ are characteristic wavenumbers of mean
and fluctuating fields, and $\epsilon_{\rm m}$ and $\epsilon_{\rm f}$
are the fractions to which these fields are helical. In the final state,
$k_{\rm m}$ will be close to the smallest wavenumber in the computational
domain, $k_1$. In the absence of shear, $\epsilon_{\rm m}$ is of order
unity, but can be less if there is shear or if the
boundary conditions do not permit fully helical
large scale fields (see below).
In the presence of shear,
$\epsilon_{\rm m}$ turns out to be inversely
proportional to the magnitude of the shear.
The value of $\tilde{k}_{\rm f}$, on the other hand, is determined
by small scale properties of the turbulence and is assumed known.

Both $k_{\rm m}$ and $k_{\rm f}$ are positive.
However, $\epsilon_{\rm m}$ can be negative which is typically the case
when $\alpha_{\rm K}<0$. The sign of $\epsilon_{\rm f}$ is defined such
that it agrees with the sign of $\epsilon_{\rm m}$, i.e.\
both change sign simultaneously and hence
$\tilde{k}_{\rm m}\tilde{k}_{\rm f}\ge0$.
In more general situations, $k_{\rm m}$ can be different from $k_1$.
Both $k_{\rm m}$ and $k_{\rm f}$ are defined more generally via
\EQ
k_{\rm m}^2=\mu_0\bra{\meanJJ\cdot\meanBB}/\bra{\meanAA\cdot\meanBB},
\label{km2}
\EN
\EQ
k_{\rm f}^2=\mu_0\bra{\jj\cdot\bb}/\bra{\aaaa\cdot\bb}.
\label{kf2}
\EN
Using \Eqs{kmtilde}{kftilde} together with \Eq{Eq41ofB01} we have
\EQ
\tilde{k}_{\rm m}\bra{\meanBB^2}=\mu_0\bra{\meanJJ\cdot\meanBB}
=-\mu_0\bra{\jj\cdot\bb}=\tilde{k}_{\rm f}\bra{\bb^2},
\label{Bfin}
\EN
which generalizes Eq.~(46) of B01 to the case with fractional helicities;
see also Eq.~(79) of BDS.

Although \Eq{Bfin} may not be precisely satisfied in simulations,
there is evidence that it is most nearly obeyed when the kinetic and
magnetic Reynolds numbers are large (B01). In the presence of shear, the large scale
magnetic energy may be time dependent, in which case \Eq{Bfin} is expected
to apply only on the time average.

Next, we want to express the final steady state values of
$\bra{\bb^2}\equiv b^2_{\rm fin}$
and $\bra{\meanBB^2}\equiv B^2_{\rm fin}$ in terms of $B_{\rm eq}^2$.
Using \Eqs{alphaKM2}{kftilde} we have first of all
\EQ
\alpha_{\rm M}=-\eta_{\rm t0}\tilde{k}_{\rm f}\bra{\bb^2}/B_{\rm eq}^2.
\label{alpM_bb2}
\EN
On the other hand, in the steady state we have from \Eqs{dABdt}{emf}
\EQ
\alpha_{\rm K}+\alpha_{\rm M}-\eta_{\rm T}\tilde{k}_{\rm m}=0,
\label{alpM_final}
\EN
where $\eta_{\rm T}=\eta+\eta_{\rm t}$ is the total
magnetic diffusivity.
These two relations yield
\EQ
{b^2_{\rm fin}\over B_{\rm eq}^2}={\alpha_{\rm K}
-\eta_{\rm T}\tilde{k}_{\rm m}\over\eta_{\rm t0}\tilde{k}_{\rm f}},\;\;\;
{B^2_{\rm fin}\over B_{\rm eq}^2}={\alpha_{\rm K}
-\eta_{\rm T}\tilde{k}_{\rm m}\over\eta_{\rm t0}\tilde{k}_{\rm m}}.
\label{final_field_strength}
\EN
In models where $\eta_{\rm t}$ is also quenched, both small scale and
large scale field strengths increase as $\eta_{\rm t}$ is more quenched.

We note that both large and small scale field saturation
strengths are determined by helicity considerations.
In case~I with $g=g_{\rm fin}$ (for the final state), we have
$\bra{\bb^2}/B_{\rm eq}^2=(C_\alpha-g_{\rm fin}
\tilde\kappa_{\rm m})\iota/\tilde\kappa_{\rm f}$.
Making use of the estimate \eq{Calp}, we have
\EQ
\bra{\bb^2}/B_{\rm eq}^2\approx
1-\iota g_{\rm fin}\tilde\kappa_{\rm m}/\tilde\kappa_{\rm f},
\EN
so $\bra{\bb^2}\approx B_{\rm eq}^2$ in the limit of large $C_\alpha$,
i.e.\ large $\tilde\kappa_{\rm f}$, and/or small $\tilde\kappa_{\rm m}$,
which is the case for $\alpha\Omega$ dynamos.
Regardless of the value of $g_{\rm fin}$, we have always
\EQ
\bra{\meanBB^2}/\bra{\bb^2}=\tilde\kappa_{\rm f}/\tilde\kappa_{\rm m}
\label{final_state}
\EN
in the final state.

We reiterate that our analysis applies to flows with helicity. In the
nonhelical case, $\alpha_{\rm K}=\tilde{k}_{\rm m}=\tilde{k}_{\rm f}=0$,
so \Eq{final_field_strength} cannot be used. Nevertheless, one would
expect a finite value of $\bra{\bb^2}$ because of small scale dynamo
action. The present approach is not really designed to handle this
case: if we multiply the first equation in (\ref{final_field_strength})
by ${\tilde k}_f$, and then set it equal 
to $0$, we would just obtain $0=0$ for that equation.

Even in the fully helical case there can be substantial
small scale contributions. Closer inspection of the runs of B01 reveals,
however, that such contributions are particularly important only
in the early kinematic phase of the dynamo.\footnote{
We should also point out that in the saturated state, $\tilde{k}_{\rm f}$
is expected to show a weak $R_{\rm m}$ dependence: assuming a Kolmogorov
spectrum for $\bb$ between $k_{\rm f}$ and the dissipation wavenumber
$k_{\rm d}$ with $k_{\rm d}/k_{\rm f}\sim R_{\rm m}^{3/4}$ one finds
$\tilde{k}_{\rm f}=k_{\rm f}R_{\rm m}^{1/4}$. During the growth phase,
however, the spectrum of $\bb$ rises with $k$ either like $k^{3/2}$
(Kulsrud \& Andersen 1992) or like $k^{1/3}$ (Brandenburg et al.\ 1996, B01),
so in either case the spectrum of $\bb$ is peaked near $k_{\rm d}$ and
$\tilde{k}_{\rm f}\approx k_{\rm d}$ may then be a better approximation.}

\subsection{Dependence of $\tilde{k}_{\rm m}$ on $\alpha$ and $S$}
\label{Sdependence}

In order to estimate $b_{\rm fin}$ and $B_{\rm fin}$, it is essential to
know the value of $\tilde{k}_{\rm m}$. We show here that for
$\alpha\Omega$ dynamos, $\tilde{k}_{\rm m}$ is inversely proportional
to the ratio of shear to turbulent velocities. The resulting relation
turns out to be well confirmed by the numerical model solutions below.

In a numerical model, the value of $\tilde{k}_{\rm m}$ can be easily
calculated for a given mean field using \Eq{kmtilde}. However, in order to
understand the dependence of $\tilde{k}_{\rm m}$ on $\alpha$ and $S$
we first consider a one-dimensional model. We  consider the case
$C_\Omega\gg C_\alpha$ (the $\alpha\Omega$ approximation), so that we
can neglect the $\alpha$ effect in the equation for the toroidal magnetic
field $B_y$. (In the opposite case, $C_\Omega\ll C_\alpha$,
we expect $\tilde{k}_{\rm m}\approx k_1$.)
We employ the gauge $\overline\phi=\meanUU\cdot\meanAA$
(Brandenburg et al.\ 1995), so \Eq{dAdt} can then be written as (BBS)
\EQ
\partial\meanAA/\partial t=-S\bar{A}_y\hat{\bf x}
+\alpha\bar{A}_x'\hat{\bf y}+\eta_{\rm T}\meanAA'',
\label{dmeanAAdt}
\EN
where primes denote $z$-derivatives, $\hat{\bf x}$ and $\hat{\bf y}$ are
unit vectors in the $x$ and $y$-directions, respectively,
and $\eta_{\rm T}=\eta+\eta_{\rm t}$ is the total magnetic diffusivity.

In a marginally excited one-dimensional model with periodic boundaries,
the solution consists of traveling waves, so all volume averages are
independent of time. In particular, $\bra{\meanJJ\cdot\meanBB}$ and
$\bra{\meanBB^2}$ are constant during a magnetic cycle. Therefore,
$\alpha$ and $\eta_{\rm T}$ are constant during the saturated state,
and we can use linear theory to find $\meanAA$ in the form
$\tilde{\bf A}e^{\ii kz-\ii\omega t}$. Real and imaginary parts of
the relevant eigenvalue (corresponding to growing solutions) are
\EQ
\omega_{\rm cyc}\equiv\mbox{Re}\,\omega
=\eta_{\rm T}k_{\rm m}^2(\tilde{C}_\alpha\tilde{C}_\Omega/2)^{1/2},
\EN
\EQ
\lambda\equiv\mbox{Im}\,\omega=\omega_{\rm cyc}-\eta_{\rm T}k_{\rm m}^2,
\EN
where $\tilde{C}_\alpha=\alpha/\eta_{\rm T}k_{\rm m}$
and $\tilde{C}_\Omega=S/\eta_{\rm T}k_{\rm m}^2$
are effective dynamo numbers based on $k_{\rm m}$
(not $\tilde{k}_{\rm m}$) during the saturated state.
The corresponding eigenvector is
\EQ
\tilde{\bf A}=-(1+\ii)
(\tilde{C}_\Omega/2\tilde{C}_\alpha)^{1/2}\hat{\bf x}
+\hat{\bf y}.
\EN
This yields for $\tilde{k}_{\rm m}=\mbox{Re}
(\tilde{\bf J}^*\tilde{\bf B})/|\tilde{\bf B}|^2$ the result
\EQ
\tilde{k}_{\rm m}=
(2\tilde{C}_\alpha\tilde{C}_\Omega)^{1/2}k_{\rm m}
/(\tilde{C}_\alpha+\tilde{C}_\Omega).
\EN
Since the saturated state corresponds to the marginally
excited state, we have $\lambda=0$, which implies
$\omega_{\rm cyc}=\eta_{\rm T}k_{\rm m}^2$ and
$\tilde{C}_\alpha\tilde{C}_\Omega=2$, so
\EQ
\epsilon_{\rm m}\equiv \tilde{k}_{\rm m}/k_{\rm m}
=2/(\tilde{C}_\alpha+\tilde{C}_\Omega).
\label{epsm_sat}
\EN
Given that we have made the $\alpha\Omega$ approximation, i.e.\
$C_\alpha\ll C_\Omega$, we have
$\epsilon_{\rm m}=2/\tilde{C}_\Omega$. 
Since $\tilde{C}_\Omega=S/\eta_{\rm T}k_{\rm m}^2$, we have
\EQ
\epsilon_{\rm m}\approx2\eta_{\rm T}k_{\rm m}^2/S.
\EN
In the marginal state, $\omega_{\rm cyc}=\eta_{\rm T}k_{\rm m}^2$,
so we can write $\epsilon_{\rm m}=2\omega_{\rm cyc}/S$, which is
useful for diagnostic purposes.
Another useful diagnostic quantity, considered also in BBS, is
the ratio of toroidal to poloidal field strength,
$Q=(\bra{\overline{B}_y^2}/\bra{\overline{B}_x^2})^{1/2}$.
In the steady state, $Q^{-1}=\epsilon_{\rm m}/\sqrt{2}$.
These relations between $Q^{-1}$, $\epsilon_{\rm m}$, and
$C_\Omega$ will later be compared with those obtained from
the one and two-dimensional models.

\subsection{The force-free degeneracy}
\label{Sdegeneracy}

Although $\alpha$ can only be obtained by solving an explicitly time dependent
differential equation, it will be of some interest to estimate the effective
values of $\alpha$ both during the growth and saturated phases, and to assess
whether or not $\alpha$ is catastrophically quenched. We first clarify
the potentially misleading finding of B01 
that the simulation results are empirically described
by a model where both $\alpha$ and $\eta_{\rm t}$ are catastrophically
quenched.  We begin by making the
assumption that the time derivative in \Eq{fullset2} can be dropped;
see Appendix~A for details. This leads to
\EQ
\alpha={\alpha_{\rm K}
+R_{\rm m}\eta_{\rm t}\mu_0\bra{\meanJJ\cdot\meanBB}/B_{\rm eq}^2
\over1+R_{\rm m}\bra{\meanBB^2}/B_{\rm eq}^2},
\label{Gruz+Diam}
\EN
which we refer to as the adiabatic approximation.
Equation \eq{Gruz+Diam} was first derived by
Gruzinov \& Diamond (1994; 1995), Bhattacharjee \& Yuan (1995)
and Kleeorin et al.\ (1995).

The adiabatic approximation can be applied
to nonoscillatory dynamos near the final steady state. For oscillatory
dynamos the adiabatic approximation is generally {\it invalid}, except in
the special case where $\bra{\meanemf\cdot\meanBB}$ is constant in time
(the one-dimensional $\alpha\Omega$ dynamos considered below are such
an example). We note that in the adiabatic approximation $k_{\rm f}$
does not enter explicitly. It only enters if one were to calculate
$\bra{\bb^2}$ for a given solution. We also note that for
an imposed uniform magnetic field we have $\meanJJ=0$, in which case
\Eq{Gruz+Diam}
predicts a catastrophically quenched $\alpha$. This is in complete agreement
with the numerical results of Cattaneo \& Hughes (1996).
One can also see how in the fully helical case $\alpha$
appears to be quenched only in a non-resistive way.

In the special case where the large scale field
is force free, which was the case in B01, we have
$\bra{\meanJJ\cdot\meanBB}\meanBB=\bra{\meanBB^2}\meanJJ$
and can then write the mean turbulent electromotive force,
$\meanemf=\alpha\meanBB-\eta_{\rm t0}\mu_0\meanJJ$, with
$\alpha$ from \Eq{Gruz+Diam} and constant $\eta_{\rm t0}$, as
\EQA
\meanemf={\alpha_{\rm K}
+R_{\rm m}\eta_{\rm t}\mu_0\bra{\meanJJ\cdot\meanBB}/B_{\rm eq}^2
\over1+R_{\rm m}\bra{\meanBB^2}/B_{\rm eq}^2}\,\meanBB
-\eta_{\rm t0}\mu_0\meanJJ
\nonumber\\
={\alpha_{\rm K}\meanBB
\over 1+R_{\rm m}\bra{\meanBB^2}/B_{\rm eq}^2}
-{\eta_{\rm t0}\mu_0\meanJJ
\over 1+R_{\rm m}\bra{\meanBB^2}/B_{\rm eq}^2}.
\label{bothquenched12}
\ENA
The reformulation allowed us
to combine the $\bra{\meanJJ\cdot\meanBB}$ term in the
$\alpha$ expression, together with the constant $\eta_{\rm t0}$ term
into a quenching expression for $\eta_{\rm t}$;
cf.\ the last term in \Eq{bothquenched12}. If one identifies
$\alpha$ with $\alpha_{\rm K}/(1+R_{\rm m}\bra{\meanBB^2}/B_{\rm eq}^2)$,
and $\eta_{\rm t}$ with
$\eta_{\rm t0}/(1+R_{\rm m}\bra{\meanBB^2}/B_{\rm eq}^2)$, then this
$\alpha$ is different from that consistent with magnetic helicity
conservation, but in this particular case it yields the same
electromotive force. This explains the excellent agreement between
the fully helical simulations of B01 and models with catastrophically
quenched $\alpha$ and $\eta_{\rm t}$. In cases with shear, for example,
this degeneracy is lifted and then only \Eq{Gruz+Diam}, or its
time dependent generalization \eq{fullset2}, can be used.

If one however does assume from the outset that $\eta_{\rm t}\propto\alpha$,
then magnetic helicity conservation and  
\Eq{Gruz+Diam} produce a slightly 
different resistively limited quenching for both
$\alpha$ and $\eta_{\rm t}$ as seen below, which is also not a bad
fit to B01 (see FB02).

Before we go on analyzing the early saturation phase we discuss in more detail
the effective value of $\alpha$ in the fully saturated state.

\subsection{The effective $\alpha$ during saturation}
\label{Seffective_alpha}

In this section we consider the value of $\alpha$ in the final saturated
state. We use the subscript `fin' for final and emphasize that the resulting
expressions are only valid in the steady state, in which case
Eqs~\eq{alpha}, \eq{alphaKM2}, and \eq{Eq41ofB01} yield
\EQ
\bra{\meanJJ\cdot\meanBB}=
-(\alpha-\alpha_{\rm K})B_{\rm eq}^2/\eta_{\rm t0}.
\EN
Therefore, as pointed out in FB02, the
$\alpha$ also appears on the right of \eq{Gruz+Diam}
and we must manipulate to solve for $\alpha$. This is given by
\EQ
\alpha = \alpha_{\rm K}{1+R_{\rm m}g_{\rm fin}\over 1+
R_{\rm m}(g_{\rm fin}+B_{\rm fin}^2/B^2_{\rm eq})},
\label{3.3}
\EN
where $g_{\rm fin}=g(B_{\rm fin})$ is the fraction by which the
turbulent magnetic diffusivity is quenched in the final state.
To determine the quenching of $\alpha$ in the steady state, 
one must specify $g(\meanBB)$ and solve for $\alpha$.  
We consider the two cases of $\eta_{\rm t}$ presented in \Sec{Sformalism}.

For case I, we can see immediately from \eq{3.3} that for large $R_{\rm m}$,
$\alpha/\alpha_{\rm K}=(1+g_{\rm fin}^{-1}B^2_{\rm fin}/B^2_{\rm eq})^{-1}$,
so $\alpha$ is not resistively quenched unless $g_{\rm fin}$ itself is
quenched to resistively small values.  
For case II, $g=\alpha/\alpha_{\rm K}$, so
we must manipulate \eq{3.3} further and find
a quadratic equation for $\alpha$. 
The appropriate solution for $R_{\rm m}\gg1$ is
\EQA
\alpha=\left\{
\begin{array}{ll}
\alpha_{\rm K}(1-B_{\rm fin}^2/B_{\rm eq}^2)&
\mbox{for $B^2_{\rm fin}<B_{\rm eq}^2$},\\
0&\mbox{otherwise},
\end{array}\right.
\label{3.4}
\ENA
so $\alpha\propto\eta_{\rm t}$ is quenched resistively
all the way to zero if $\bra{\meanBB^2}/B_{\rm eq}^2>1$,
which is usually the case in the simulations.

Using the assumption that a nontrivial stationary state is reached
at late times (justified by simulations)
several important points  are revealed by the above results.
First, the reason  $\alpha$ can only be weakly quenched for an unquenched
$\eta_{\rm t}$ is that otherwise the field would eventually decay through 
the action of $\eta_{\rm t}$, precluding a stationary state in the first place.
\Eq{3.3} also shows that 
$\alpha$ is quenched more strongly than $\eta_{\rm t}$ when 
$\eta_{\rm t}$ is  independent of $\alpha$, 
whereas if $\eta_{\rm t}\propto\alpha$ then both $\alpha$ and $\eta$ 
are quenched in tandem.
We therefore expect a higher saturation value of the field strength
for case II of \Sec{Sformalism} as compared to case I with $\tilde{g}=0$.

Finally, in the case $\eta_{\rm t} \propto \alpha$ (case II)
the fact that the saturation ratio of the mean field to the equipartition value
turns out to be $\gg\alpha/\alpha_{\rm K}$ at saturation
is important because it suggests that the 
early time growth must have a less resistively limited form of $\alpha$.
We discuss this in the next section.  
In this context, again note how one might be misled
by uniform field simulations designed to measure $\alpha$, such
as those of Cattaneo \& Hughes (1994), in which the mean field cannot grow.

\subsection{Early time evolution}
\label{Searly}

During the early growth phase
the magnetic helicity varies on time scales shorter than
the resistive time, so
the last term in \Eq{fullset2} can be neglected and so $\alpha_{\rm M}$
evolves then approximately according to
\EQ
{\dd\alpha_{\rm M}\over\dd t}\approx-2\eta_{\rm t0}k_{\rm f}^2
(\alpha_{\rm K}+\alpha_{\rm M}-\eta_{\rm t0}\tilde{k}_{\rm m})
{\bra{\meanBB^2}\over B_{\rm eq}^2}.
\label{dynquench_early}
\EN
We use this equation
to describe the early kinematic time evolution when $\bra{\meanBB^2}$, and hence
also $\alpha_{\rm M}$, grow exponentially.  FB02 showed
that the early time evolution leads to a nearly $R_{\rm m}$ independent
growth phase at the end of which a significant large scale field growth
occurs. We now derive this in our present formalism.

In the following discussion we restrict ourselves to the case
where $\eta$ is small. Magnetic helicity conservation then requires that
\EQ
\bra{\meanAA\cdot\meanBB}\approx-\bra{\aaaa\cdot\bb}\quad
\mbox{(for $t\leq t_{\rm kin}$)},
\label{helequil}
\EN
where the time $t=t_{\rm kin}$ marks the end of the exponential
growth phase (and the `initial' saturation time $t_{\rm sat}$ used in B01).
This time is determined by the condition that the
term in brackets in \Eq{dynquench_early} becomes significantly reduced,
i.e.\ $-\alpha_{\rm M}$
becomes comparable to $\alpha_{\rm K}-\eta_{\rm t0}\tilde{k}_{\rm m}$.
Using \eq{alpM_bb2} and \eq{helequil} together with
\eqs{kmtilde}{km2}, we obtain
the mean squared field strengths of the small and large scale fields
at the end of the kinematic phase,
\EQ
{b_{\rm kin}^2\over B_{\rm eq}^2}={\alpha_{\rm K}
-\eta_{\rm t0}\tilde{k}_{\rm m}\over\tilde{\iota}\eta_{\rm t0}\tilde{k}_{\rm f}},\;\;\;
{B_{\rm kin}^2\over B_{\rm eq}^2}={\alpha_{\rm K}
-\eta_{\rm t0}\tilde{k}_{\rm m}\over\tilde{\iota}\eta_{\rm t0}\tilde{k}_{\rm m}}
{k_{\rm m}^2\over k_{\rm f}^2},
\label{kinematic_field_strength}
\EN
respectively, where we have included the extra correction factor
\EQ
\tilde{\iota}=1+R_{\rm m}^{-1}
{k_{\rm f}/\epsilon_{\rm f}\over k_{\rm m}/\epsilon_{\rm m}},
\EN
which becomes important for intermediate values of $R_{\rm m}$.
This correction factor results from restoring the $\alpha_{\rm M}/R_{\rm m}$
term from \Eq{fullset2} in \Eq{dynquench_early}.
Not surprisingly, at the end of the kinematic phase the
small scale magnetic energy is almost the same as in the final state;
cf.\ \Eq{final_field_strength}. However, the large scale magnetic energy is
still by a factor $k_{\rm m}^2/k_{\rm f}^2$ smaller than in the final state
(although $\epsilon_{\rm m}$ may be somewhat different in the two stages).
This result was also obtained by Subramanian (2002) using a similar approach.
Using \Eqs{Calp}{kinematic_field_strength}, we can write
\EQ
{B_{\rm kin}^2\over B_{\rm eq}^2}=
{k_{\rm m}/\epsilon_{\rm m}\over\tilde{\iota}k_{\rm f}/\epsilon_{\rm f}}
\left(1-{\tilde{\kappa}_{\rm m}\over\tilde{\kappa}_{\rm f}}\right),
\label{kinematic_field_strength2}
\EN
which shows that $B_{\rm kin}$ can be comparable to and even
in excess of $B_{\rm eq}$, especially when
$\epsilon_{\rm m}$ is small (strong shear).

As emphasized in FB02, for an $\alpha^2$ dynamo, the initial evolution to $B_{\rm kin}$
is significantly more optimistic an estimate than what could have
been expected based on lorentzian quenching.
In the case of an $\alpha\Omega$ dynamo, $\tilde{k}_{\rm m}\ll k_{\rm m}$,
so $B_{\rm kin}$ can be correspondingly larger. In fact, for
\EQ
\epsilon_{\rm m}/\epsilon_{\rm f}\leq k_{\rm m}/k_{\rm f},
\EN
the large scale field begins to exceed the small scale field already during
the kinematic growth phase. Using the estimate
$\epsilon_{\rm m}\approx2/\tilde{C}_\Omega$ (\Sec{Sdependence}),
and since $\tilde{C}_\Omega=C_\Omega$ during the kinematic stage,
we see that large and small scale fields become comparable when
$\epsilon_{\rm f}C_\Omega\geq2k_{\rm f}/k_{\rm m}$.
Combining this with the condition for a marginally excited dynamo,
$C_\alpha C_\Omega=2$, we have
$\epsilon_{\rm f}C_\Omega\geq2(\iota/\kappa_{\rm m})^{1/2}\approx2$.
In simulations of rotating convection, $\epsilon_{\rm f}\approx0.03$
(Brandenburg et al.\ 1996); assuming that this relatively low value
of $\epsilon_{\rm f}$ is generally valid, we have
$C_\Omega\ga\epsilon_{\rm f}/2\approx60$
for the condition above which the large scale field exceeds the
small scale field already during the kinematic growth phase. This
condition is likely to be satisfied both for stellar and galactic dynamos.

During the subsequent resistively limited saturation phase the energy of
the large scale field grows first linearly,
\EQ
\bra{\meanBB^2}\approx B_{\rm kin}^2+2\eta k_{\rm m}^2(t-t_{\rm kin})
\quad\mbox{(for $t>t_{\rm kin}$)},
\label{pastkin}
\EN
and saturates later in a resistively limited fashion; see Eq.~(45) of B01.

In the limit of large $R_{\rm m}$ for  the maximally helical case,
(\ref{helequil}) implies that 
$\bra{\meanBB^2}$ remains of the order
of $B^2_{\rm kin}$ for times  $t_{\rm kin}< t\ll1/\eta k_{\rm f}^2$.
Using (\ref{alpha}), (\ref{alphaKM}), and (\ref{ab_to_alpM})
we also find that toward the end of the kinematic regime, $\alpha$ is
quenched non-resistively as
\EQ
\alpha/\alpha_{\rm K}\approx1-\bra{\meanBB^2}/B^2_{\rm kin}
\quad\mbox{(for $t\approx t_{\rm kin}$}),
\EN
where we have assumed $\alpha_{\rm K}\gg\eta_{\rm t0}\tilde{k}_{\rm m}$.
The fact that this is independent of $R_{\rm m}$ and of the
choice of $\eta_{\rm t}$, contrasts with the $\alpha$ formulae for the 
late-time regime considered in the previous
section. This highlights the need for a 
fully time dependent dynamical theory
to understand the time dependence of $\alpha$ quenching.

In the following section we present and discuss 
results from simple mean-field models
using dynamical quenching.

\section{The full time evolution}
\label{Sfull}

We will first consider a one-dimensional $\alpha^2\Omega$ dynamo model
with constant shear, $\meanUU=(0,Sx,0)$.
Such a model is sometimes used to model stellar dynamo
waves traveling in the latitudinal direction (e.g.\ Robinson \& Durney 1982),
where $(x,y,z)$ are identified with spherical coordinates $(r,\phi,-\theta)$.
In terms of the mean
vector potential $\meanAA(z,t)$, the uncurled mean-field induction equation
reads (BBS)
\EQ
\partial\meanAA/\partial t=\meanemf-S\bar{A}_y\hat{\bf x}
-\eta\mu_0\meanJJ,
\label{dmeanAAdt2}
\EN
where $\mu_0\meanJJ=-\partial^2\meanAA/\partial z^2$ and $\overline{A}_z=0$. In contrast to
\Eq{dmeanAAdt} we do not make the $\alpha\Omega$ approximation.
Later we also consider two-dimensional models which can be compared
with simulations. For quick parameter surveys, however, the one-dimensional
models in the one-mode truncation are quite useful.

\subsection{The one-mode truncation}
\label{Sone-mode}

We first consider
the one-mode truncation ($k=k_{\rm m}=k_1$), i.e.\ we assume
$\meanAA=\hat{\AAA}e^{\ii k_{\rm m} z}$, where $\hat{\AAA}(t)$ is complex, 
and solve the set of two ordinary differential equations
for $\hat{A}_x$ and $\hat{A}_y$,
\EQ
\dd\hat{\AAA}/\dd t
=\hat{\emf}-S\hat{A}_y\hat{\bf x}-\eta\mu_0\hat{\JJ},
\label{dhatAAdt}
\EN
where $\mu_0\hat{\JJ}=k_{\rm m}^2\hat{\AAA}$. The two components of the magnetic
field are $\hat{B}_x=-\ii k_{\rm m}\hat{A}_y$ and $\hat{B}_y=\ii k_{\rm m}\hat{A}_x$. The
electromotive force is $\hat{\emf}=\alpha\hat{\BB}-\eta_{\rm t}\mu_0\hat{\JJ}$,
where $\alpha$ is given by \Eq{alpha} and $\alpha_{\rm M}$ is obtained
by solving \Eq{dynquench} using
$\bra{\emf\cdot\BB}=\mbox{Re}(\hat{\emf}^*\cdot\hat{\BB})$,
where asterisks denote complex conjugation.
For diagnostic purposes we also monitor
$\tilde{k}_{\rm m}=\mbox{Re}(\hat{\JJ}^*\cdot\hat{\BB})/|\BB|^2$.

\begin{figure}[t!]\centering\includegraphics[width=0.45\textwidth]{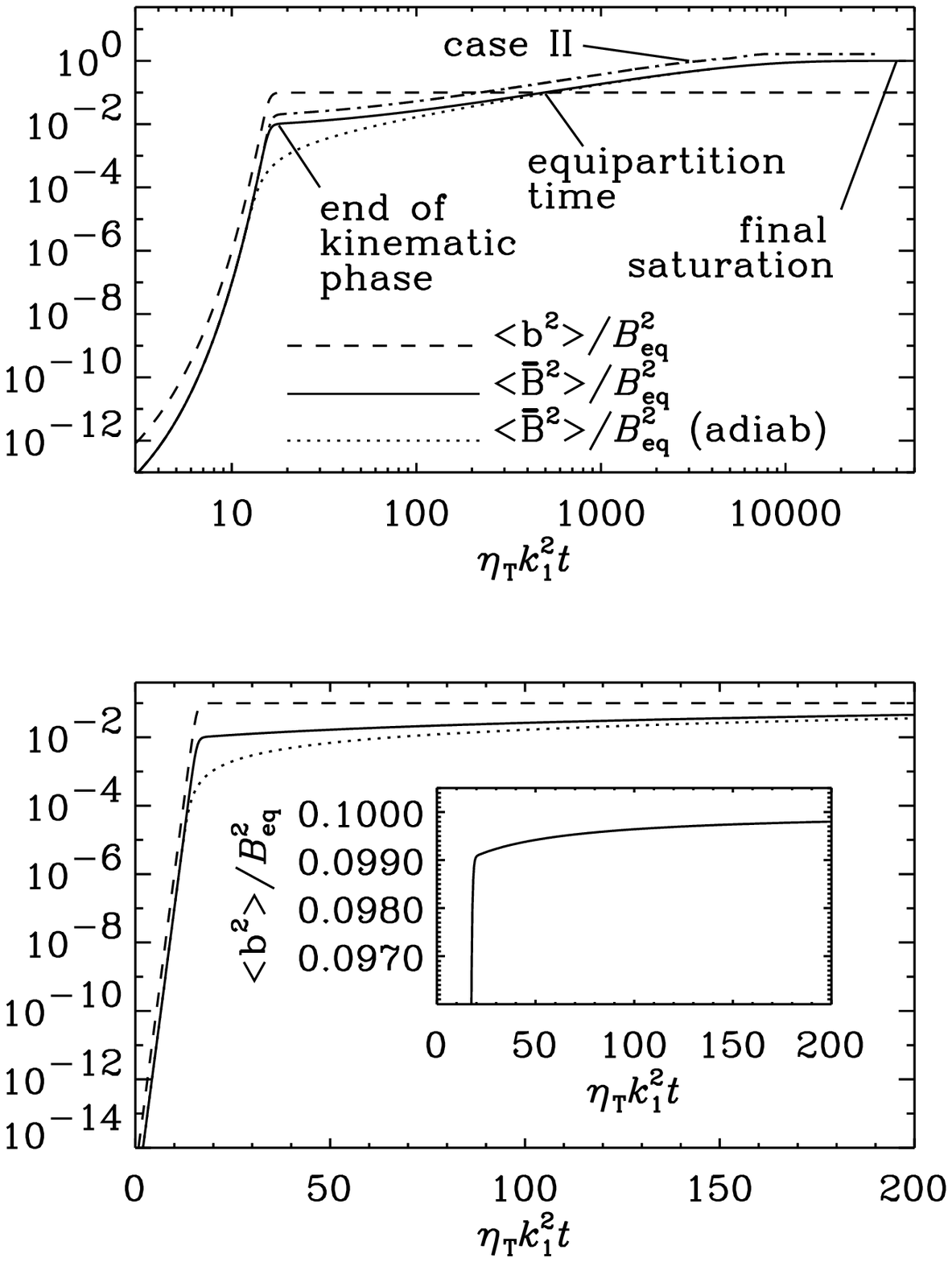}\caption{
%\begin{figure}[t!]\plotone{f1.eps}\caption{
The upper panel shows the evolution of the dimensionless 
$\bra{\meanBB^2}$ and $\bra{\bb^2}$ (solid and dashed lines, respectively,
and both in units of $B_{\rm eq}^2$)
in a doubly-logarithmic plot for an $\alpha^2$ dynamo with
%weakly quenched $\eta$ from 
%AB: note, we have g-tilde=0!
$\eta_{\rm t}=\mbox{const}$ 
(case~I).  The adiabatic approximation
gives significantly smaller values of $\bra{\meanBB^2}$ by the end
of the kinematic growth phase.
The dash-dotted line gives $\bra{\meanBB^2}$ for case II
($\eta_{\rm t}\propto\alpha$); $\bra{\bb^2}$ is essentially the
same in cases I and II.
Note that time is displayed on a logarithmic scale to
accommodate comfortably both the kinematic phase and the
final saturation.
The lower panel gives the same three
curves in a semi-logarithmic plot. The inset shows the slow and
small ($\sim1\%$) adjustment of $\bra{\bb^2}$ during the early saturation phase.
$R_{\rm m}=10^4$, $C_\alpha=2$, $C_\Omega=0$, $\kappa_{\rm f}=10$,
$\epsilon_{\rm f}=1$, $\tilde{g}=0$.
}\label{Fpcrossing}\end{figure}

In \Fig{Fpcrossing} we plot the evolution of $\bra{\meanBB^2}$ and $\bra{\bb^2}$
for $R_{\rm m}=10^4$, $C_\alpha=2$, $k_{\rm f}=5$, $\epsilon_{\rm f}=1$. 
Initially, both
quantities grow exponentially at the rate $\lambda\equiv\alpha_{\rm K}k_1-\eta_{\rm T0}k_1^2$.
This phase stops rather abruptly at $t_{\rm kin}=\lambda^{-1}\ln(B_{\rm kin}/B_{\rm ini})$,
and turns then into a resistively limited growth phase.
Note, however, that already by the end of the kinematic growth phase
the large scale field is a certain fraction of the equipartition field
strength, which is independent of the magnetic Reynolds number.

The equipartition time can be obtained by setting
$\bra{\meanBB^2}=b_{\rm kin}^2$ in \Eq{pastkin} and
using \Eq{kinematic_field_strength}, so
\EQ
2\eta k_{\rm m}^2(t-t_{\rm kin})=(1-k_{\rm m}/k_{\rm f})^2,
\label{sattime}
\EN
where we have used, for simplicity, $\epsilon_{\rm m}=\epsilon_{\rm f}=1$.
Otherwise the expression on the right hand side of \Eq{sattime} would be
more complicated. The main point was to show that the equipartition time
is still resistively limited, which is consistent with \Fig{Fpcrossing}.

In the example shown in \Fig{Fpcrossing}, $C_\alpha=2$, so the dynamo
is only weakly supercritical and the final field strengths are
$b^2_{\rm fin}
\approx0.1\times B_{\rm eq}^2$ and $\bra{\meanBB^2}\approx B_{\rm eq}^2$,
in agreement with \Eq{final_field_strength}.

\begin{figure}[t!]\centering\includegraphics[width=0.45\textwidth]{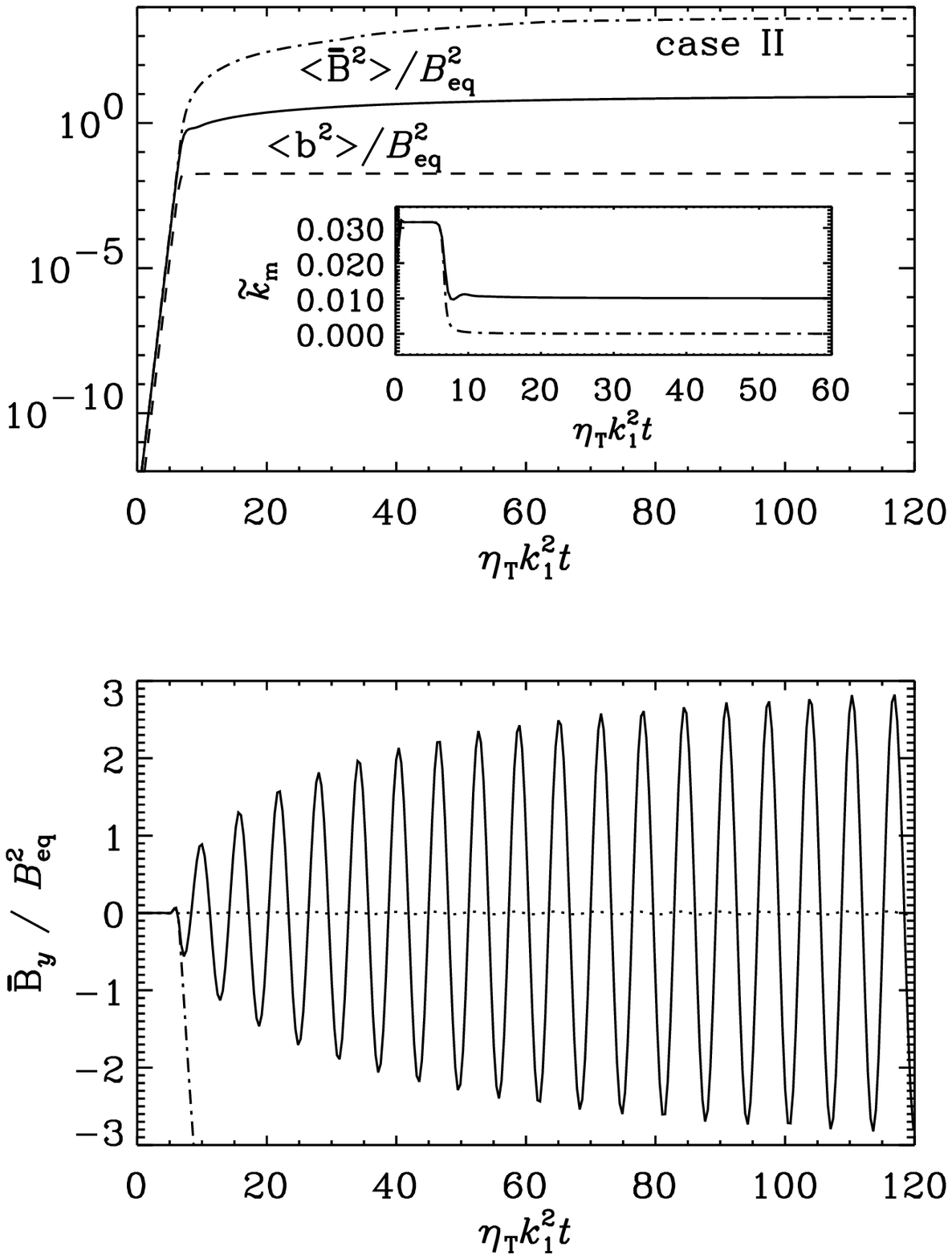}\caption{
%\begin{figure}[t!]\plotone{f2.eps}\caption{
The upper panel shows the evolution of the dimensionless
$\bra{\meanBB^2}$ and $\bra{\bb^2}$ for an $\alpha^2\Omega$ dynamo
(solid and dashed lines, respectively).
The dash-dotted line gives $\bra{\meanBB^2}$ for case II
($\eta_{\rm t}\propto\alpha$); $\bra{\bb^2}$ is essentially the
same in cases I and II.
The inset shows the evolution of
$\tilde{k}_{\rm m}$. Note that in case~II, $\tilde{k}_{\rm m}\approx0$.
The lower panel shows ${\overline B}_y$ (solid line)
and ${\overline B}_x$ (dashed line). The cycle frequency is of order
$\eta_{\rm T}k_1^2$, but the cycle amplitude adjusts on a resistive time scale.
$R_{\rm m}=10^2$, $C_\alpha=0.1$, $C_\Omega=200$, $\kappa_{\rm f}=5$,
$\epsilon_{\rm f}=1$, $\tilde{g}=0$.
In case~II (dash-dotted line) the dynamo period is very long
and the amplitude much higher.
}\label{Fpcrossing_ao}\end{figure}

We recall that in the fully helical case, because of the force-free
degeneracy, the adiabatic approximation coincides with the catastrophic
quenching hypothesis (\Sec{Sdegeneracy}), and it reproduces the final
saturation phase rather well (B01). For larger values of $R_{\rm m}$,
however, the adiabatic approximation gives significantly lower values
of $\bra{\meanBB^2}$ by the end of the kinematic growth phase (see the
dotted line in \Fig{Fpcrossing}). This difference increases with
increasing values of $R_{\rm m}$.

In the case of an $\alpha^2\Omega$ dynamo, the overall evolution of small and
large scale magnetic energy is similar, except that the large scale field is in
general not fully helical ($\epsilon_{\rm m}\ll1$), because the toroidal magnetic
field can be amplified regardless of magnetic helicity. It only relies on the presence of
a small poloidal field which must still be regenerated by the $\alpha$ effect.
The final field amplitude can be much larger than for the $\alpha^2$
dynamo. In particular, as pointed out in \Sec{Searly}, the mean field can
exceed the small scale field already during the entire kinematic growth phase if
$C_\Omega$ is large enough. This is the case in the example depicted in
\Fig{Fpcrossing_ao}, so there is no crossing of the two curves as in \Fig{Fpcrossing}.

For case~II ($\eta_{\rm t}\propto\alpha$), the saturation field strength
of the large scale field is significantly enhanced. Also, because $\eta_{\rm T}$
is now much lower in the saturated state, the value of $\epsilon_{\rm m}$
(and hence $\tilde{k}_{\rm m}$) is now strongly suppressed. This is
consistent with \Eq{epsm_sat}. Furthermore, $\eta_{\rm t}$ is now
suppressed down to the microscopic value, so the dynamo period has
increased by a factor $R_{\rm m}$.

\subsection{Comparison with B01}

In the simulations of B01, Runs~1--3 had a magnetic Prandtl number of
unity ($\nu/\eta=1$) and $R_{\rm m}\approx u_{\rm rms}/\eta k_{\rm f}$
varied from 2.4 to 18. The run with the highest magnetic Reynolds number
was Run~5 with $R_{\rm m}\approx100$, but here $\nu/\eta=100$, so the
hydrodynamic Reynolds number was unity and there was basically no turbulent mixing.
Shear was absent in those runs, so $C_{\Omega}=0$.
The other dynamo parameters can be estimated as
$C_{\alpha}\approx\tilde{\kappa}_{\rm f}/\iota\approx5$; see \Eq{Calp}.
Assuming $\tilde{g}=0$, i.e.\ $\eta_{\rm T}=\eta_{\rm T0}$,
we have from \Eq{final_field_strength}
for the final mean squared field strength
\EQ
B_{\rm fin}^2/B_{\rm eq}^2
=\tilde\kappa_{\rm f}/\tilde\kappa_{\rm m}-\iota=3.6\dots4.0,
\EN
for Runs~1--3.
Here we have put $\tilde{\kappa}_{\rm m}=1$ for the effective
nondimensional wavenumber of the large scale field.
The actual values of $B_{\rm fin}^2/B_{\rm eq}^2$
are somewhat smaller (for Run~3, for example,
$B_{\rm fin}^2/B_{\rm eq}^2=3.6$ instead of 3.9).
The agreement with the theoretically expected value is quite
reasonable even with $\tilde{g}=0$. Full agreement could
in principle be achieved with negative values of $\tilde{g}$
($\tilde{g}=-0.13$ for Run~3, for example).
It is more likely, however, that the actual value of $C_\alpha$ is
somewhat less than our estimate $\tilde{\kappa}_{\rm f}/\iota$.

\begin{figure}[t!]\centering\includegraphics[width=0.45\textwidth]{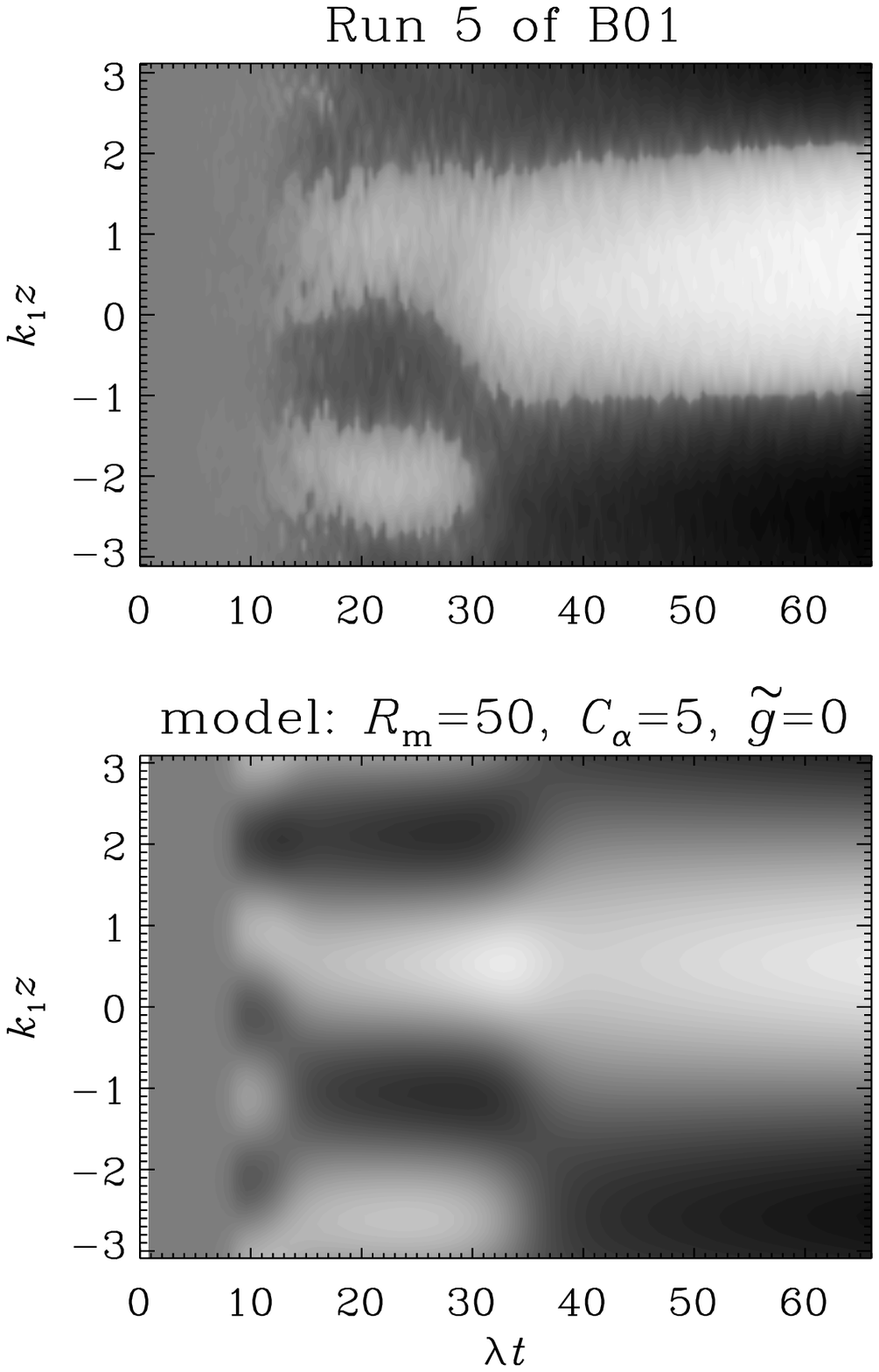}\caption{
%\begin{figure}[t!]\plotone{f3.eps}\caption{
Comparison of the space-time (or butterfly) diagram from Run~5 of B01 with
that from the one-dimensional model.
Dark (light) shades indicate negative (positive) values.
$R_{\rm m}=50$, ${\cal D}=5$, $\tilde{g}=0$.
}\label{Fpbutter_run5}\end{figure}

The saturation behavior of $\bra{\meanBB^2}(t)$, as seen in the
simulations of B01, was already well reproduced by the adiabatic
approximation (see Fig.~21 of B01). This is because the values of
$R_{\rm m}$ are still too small to be able to see significant differences between
dynamical quenching and the adiabatic approximation.
For Run~5 there is however a noticeable slow-down in the saturation
behavior of the field at $\lambda t\approx15$ if the large
scale field is identified with the spectral energy at $k=k_1$.
This behavior is well reproduced by the one-dimensional model
if $R_{\rm m}=50$ is chosen. The nominal value of $R_{\rm m}$
is actually around 100, but this is probably unrealistic and
would also give too high values of the kinematic growth rate
compared with the simulation. The slow-down at $\lambda t\approx15$,
which is also seen in the one-dimensional model, is not reproduced
in the one-mode truncation. The reason for this difference lies in the fact
that at early and intermediate times, $k_{\rm m}=2$ prevails, and only
at later times $k_{\rm m}=1$ becomes dominant; see \Fig{Fpbutter_run5}.

In order to test the dynamical quenching theory more quantitatively, it
would be useful to produce new simulations with smaller scale separation,
e.g.\ $k_{\rm f}=2\ldots3$, and an initial seed magnetic field where
only one of several possible large scale eigenfunctions are present.

\subsection{$\alpha^2\Omega$ dynamos: a parameter study}
\label{Salpom}

\begin{table*}[t!]
\caption{The effect of changing $C_\alpha$ and $C_\Omega$
in an $\alpha^2\Omega$ dynamo with dynamical $\alpha$ quenching
using the one-mode approximation. For all runs, $R_{\rm m}=20$, $\tilde{g}=0$.
}\label{Tsum_Calp}
\begin{tabular}{rcrccccccl}
$C_\alpha$ & $C_\Omega$ & $S/\eta k_1^2$
& $b^2_{\rm fin}/B_{\rm eq}^2$
& $B_{\rm fin}^2/B_{\rm eq}^2$
& $Q^{-1}$
& $\epsilon_{\rm m}$ 
& $\omega_{\rm cyc}/S$
& $\lambda/S$
& remark \\
\vspace{-3mm}
\\
\hline
3.0&  0 &   0 & 0.42 &  2.1 & 1.00  & 1.00  &   0   &$2.0/C_\Omega$& $\alpha^2$ dynamo\\
3.0&  2 &  42 & 0.48 &  3.4 & 0.58  & 0.71  & 0.36  &  1.1         & $C_\alpha/C_\Omega={\cal O}(1)$\\
3.0& 20 & 420 & 0.59 &  16  & 0.11  & 0.12  & 0.079 &  0.27        & larger $C_\Omega$\\
1.0& 20 & 420 & 0.19 &  9.5 & 0.071 & 0.10  & 0.050 &  0.12        & smaller $C_\alpha$\\
0.3& 20 & 420 & 0.042&  2.1 & 0.071 & 0.10  & 0.050 &  0.045       & smaller $C_\alpha$\\
0.1& 20 & 420 &  0   &   0  & 0.071 & 0.100 & 0.050 &   0          & marginal case\\
0.1&200 & 4200 & 0.019&  9.5 & 0.007 & 0.010 & 0.005 &  0.012      & same $C_\alpha C_\Omega$ as in line 3\\
\end{tabular}
\end{table*}

When shear is included ($C_\Omega\neq0$), toroidal field can be
regenerated solely by the $S$-term in \Eq{dhatAAdt}.
When $C_\Omega\gg C_\alpha$, the dynamo efficiency is governed
by the product $C_\alpha C_\Omega$.
This is the regime where dynamical and fixed-form 
algebraic quenching lead to very different behaviors.

In order to study the effect of changing various input parameters
we begin with \Tab{Tsum_Calp} where we show the results for different
values of $C_\alpha$. Note that only for small values of $C_\alpha$
do the results for $\epsilon_{\rm m}$ and $\omega_{\rm cyc}/S$ agree
with the prediction of \Sec{Sdependence}.
This is simply because the $\alpha\Omega$ approximation made in
\Sec{Sdependence} is only valid for $C_\alpha\ll C_\Omega$.
As $C_\alpha$ increases,
the field strength increases approximately as predicted by
\Eq{final_field_strength}.

Increasing the value of $R_{\rm m}$ has no effect on the field geometry
and time scales ($Q^{-1}$, $\epsilon_{\rm m}$, and
$\lambda$ are unaffected); see \Tab{Tsum_Rm}.
The field strength changes only when $R_{\rm m}$ is close to unity.
Once $R_{\rm m}$ is above a certain value, the results are
essentially independent of $R_{\rm m}$.
For $\tilde{g}>0$, the field strength generally
increases, as expected, and the cycle frequency decreases;
see \Tab{Tsum_g}. One can also verify that $\epsilon_{\rm m}$ decreases
as $\tilde{g}$ is increased.
Models where $\eta_{\rm t}\propto\alpha$ (case~II) tend to produce
long cycle periods if the dynamo is sufficiently supercritical; see
\Tab{Tsum_caseII}.

\begin{table}[t!]
\caption{Effect of changing $R_{\rm m}$ for two different values of
$C_\alpha$ in the one-mode approximation.
The increase of small and large scale field strength for small values
of $R_{\rm m}$ is explained by the large values of $\iota$.
For all the models with $C_\alpha=1.0$ we find
$Q^{-1}=0.021$, $\epsilon_{\rm m}=0.014$,
$\omega_{\rm cyc}/S=0.015$, $\lambda/S=0.045$, whilst
for all the models with $C_\alpha=0.1$ we find
$Q^{-1}=0.005$, $\epsilon_{\rm m}=0.007$,
$\omega_{\rm cyc}/S=0.003$, $\lambda/S=0.011$.
For all runs, $C_\Omega=300$, $\tilde{g}=0$, and
$S/\eta k_1^2\equiv\iota R_{\rm m}C_\Omega$
varies between $6\times10^2$ and $3\times10^5$.
}\label{Tsum_Rm}
\begin{tabular}{rccc}
$R_{\rm m}$ & $C_\alpha$
& $b^2_{\rm fin}/B_{\rm eq}^2$
& $B^2_{\rm fin}/B_{\rm eq}^2$\\
\vspace{-3mm}
\\
\hline
1000& 1.0 & 0.19 & 15 \\
100 & 1.0 & 0.19 & 15 \\
 10 & 1.0 & 0.21 & 17 \\
  1 & 1.0 & 0.38 & 30 \\
100 & 0.1 & 0.019& 13 \\
 10 & 0.1 & 0.021& 15 \\
  1 & 0.1 & 0.037& 28 \\
\end{tabular}
\end{table}

\begin{table}[t!]
\caption{Effect of changing $\tilde{g}$ in the one-mode approximation.
For all runs, $R_{\rm m}=20$, $C_\alpha=0.1$,
$C_\Omega=200$, $S/\eta k_1^2=4200$,
$b^2_{\rm fin}/B_{\rm eq}^2=0.02$,
$\lambda/S=0.012$.
}\label{Tsum_g}
\begin{tabular}{rcrcccccccc}
$\tilde{g}$
& $B_{\rm fin}^2/B_{\rm eq}^2$
& $Q^{-1}$
& $\epsilon_{\rm m}$ 
& $\omega_{\rm cyc}/S$ \\
\vspace{-3mm}
\\
\hline
 0  & 9.5 & 0.007 & 0.010 & 0.0050\\
0.3 & 24  & 0.0031& 0.0043& 0.0021\\
1.0 & 70  & 0.0011& 0.0015& 0.0010\\
\end{tabular}
\end{table}

\begin{table}[t!]
\caption{Effect of changing $C_\Omega$ for case~II in the one-mode approximation.
The critical value for dynamo action is $C_\Omega=20$.
For all runs, $R_{\rm m}=20$, $C_\alpha=0.1$,
$b^2_{\rm fin}/B_{\rm eq}^2=0.02$,
}\label{Tsum_caseII}
\begin{tabular}{rcrcccccccc}
$C_\Omega$
& $B_{\rm fin}^2/B_{\rm eq}^2$
& $Q^{-1}$
& $\epsilon_{\rm m}$ 
& $\omega_{\rm cyc}/S$ \\
\vspace{-3mm}
\\
\hline
 22 & 31 & 0.0032& 0.0045& 0.0023\\
 30 & 43 & 0.0023& 0.0033& 0.0017\\
 50 & 75 & 0.0014& 0.0019& 0.0010\\
100 &161 & 0.0007& 0.0010& 0.0005\\
\end{tabular}
\end{table}

\subsection{A two-dimensional $\alpha^2\Omega$ dynamo}
\label{Salpom_2D}

In order to compare with the simulations of BBS, it is important to
consider the appropriate geometry and shear profile. As in BBS we use
sinusoidal shear, $\meanUU=(0,Sk_1^{-1}\cos k_1 x,0)$, and the mean field
is $\meanBB=\meanBB(x,z,t)$; cf.\ \Sec{Scomplete}. The results are shown
in \Tab{Tsum_2D}. The calculations have been carried out using a sixth
order finite difference scheme in space and a third order Runge-Kutta
scheme in time.

\begin{table*}[t!]
\caption{Results from the two-dimensional
$\alpha^2\Omega$ dynamo with dynamical $\alpha$ quenching.
In the last two rows, the results from the simulations of BBS and BDS
are given for comparison. Models AG2 and perhaps also R1 show some
tentative agreement with BBS; the corresponding numbers are shown
in bold face.
Models~s3 and S1 give some tentative agreement with BDS (where
$S/\eta k_1^2=1000$); the corresponding numbers are shown in italics.
}\label{Tsum_2D}
\begin{tabular}{cccrcccccccc}
&$R_{\rm m}$ & $C_\alpha$ & $C_\Omega$ & $\tilde{g}$ & $S/\eta k_1^2$
& $b^2_{\rm fin}/B_{\rm eq}^2$
& $B^2_{\rm fin}/B_{\rm eq}^2$
& $Q^{-1}$
& $\epsilon_{\rm m}$ 
& $\omega_{\rm cyc}/S$
& $\lambda/S$ \\
\hline
A1&20 &  0.3&100 & 0 &2000 & 0.05 &   4.0& 0.032 & 0.068 & 0.015 &  0.018\\ %(alp32)
A2&20 &  1.0&100 & 0 &2000 & 0.20 &  15  & 0.031 & 0.065 & 0.016 &  0.046\\ %(alp23)
A3&20 &  3.0&100 & 0 &2000 & 0.62 &  48  & 0.031 & 0.064 & 0.015 &  0.098\\ %(alp33)
\hline
{\bf R1}
  &20 &  1.0&100 & 0 &2000 & 0.20 &{\bf15}&0.031&{\bf0.065}&0.016&  0.044\\ %(alp23, alp41)
R2&50 &  1.0& 40 & 0 &2000 & 0.17 &  5.5 & 0.076 & 0.16  & 0.035 &  0.055\\ %(alp42)
R3&100&  1.0& 20 & 0 &2000 & 0.14 &  2.3 & 0.15  & 0.30  & 0.072 &  0.061\\ %(alp40)
\hline
G1&20 &  0.3&100 & 0 &2000 & 0.05 &   4.0& 0.032 & 0.068 & 0.015 &  0.018\\ %(alp32)
G2&20 &  0.3&100 &0.1&2000 & 0.05 &   5.2& 0.025 & 0.053 & 0.011 &  0.018\\ %(alp34)
G3&20 &  0.3&100 &0.3&2000 & 0.06 &   8.2& 0.018 & 0.036 & 0.009 &  0.018\\ %(alp35)
G4&20 &  0.3&100 &1.0&2000 & 0.06 &  23  & 0.007 & 0.014 & 0.007 &  0.018\\ %(alp36)
G5&20 &  0.3&100 &3.0&2000 & 0.06 &  56  & 0.003 & 0.006 & 0.0013&  0.018\\ %(alp37)
\hline
AG1&100&  0.3& 20 &3.0&2000 & 0.00 &   0  &  --   &  --   &  --   &    0  \\ %(alp43)
{\bf AG2}
   &100&  0.5& 20 &3.0&2000 & 0.10&{\bf22}&{\bf0.011}&0.024&{\bf0.006}&{\bf0.021}\\ %(alp45)
AG3&100&  1.0& 20 &3.0&2000 & 0.20 &  70  & 0.007 & 0.015 & 0.004 &  0.052\\ %(alp44)
\hline
s1& 30& 0.25& 33 &1.0&1000 & 0.05 &   3  & 0.035 & 0.072 & 0.017 &  0.007\\ %(alp53)
s2& 30& 0.30& 33 &1.0&1000 & 0.06 &   4  & 0.032 & 0.066 & 0.015 &  0.012\\ %(alp52)
{\it s3}
  & 30& 0.35& 33 &1.0&1000 & 0.07 &   6  & 0.029&{\it0.061}&{\it0.014}&{\it0.016}\\ %(alp54)
{\it S1}
  & 30& 0.35& 33 &3.0&1000 & 0.07&{\it19}& 0.009 & 0.019 & 0.005 &  0.016\\ %(alp51)
S2& 50& 0.35& 20 &3.0&1000 & 0.07 &  10  & 0.017 & 0.035 & 0.008 &  0.006\\ %(alp50)
S3& 50&  0.4& 20 &3.0&1000 & 0.08 &  14  & 0.015 & 0.031 & 0.006 &  0.011\\ %(alp49)
S4&100&  0.4& 20 &3.0&2000 & 0.08 &  14  & 0.014 & 0.028 & 0.008 &  0.011\\ %(alp48)
\hline
BDS&$\sim30$&1--2&--&--&1000&  4   & 20  & 0.018 & 0.11 &0.013--0.015&0.006\\%(BDS)
BBS&$\sim80$&1--2&--&--&2000&  6   & 30  & 0.014 & 0.06 &0.005--0.010&0.015\\%(BBS)
\end{tabular}
\end{table*}

\begin{figure}[t!]\centering\includegraphics[width=0.45\textwidth]{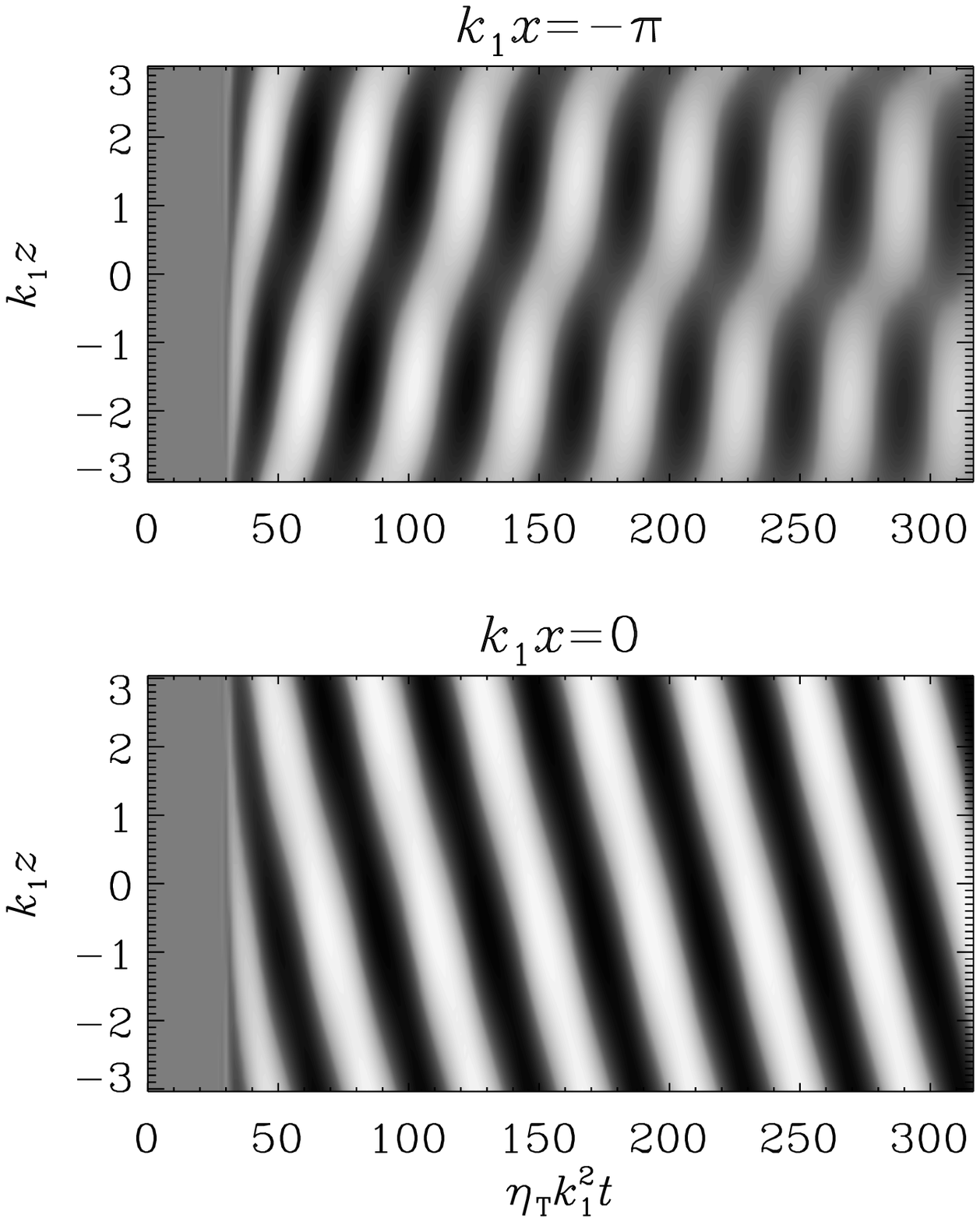}\caption{
%\begin{figure}[t!]\plotone{f4.eps}\caption{
Space-time (or butterfly) diagram of ${\overline B}_y$ for
Model~S1 with dynamical quenching. At $x=-\pi$ ($x=0$) the
shear has attained and negative (positive) maximum.
Dark (light) shades indicate negative (positive) values.
$R_{\rm m}=30$, $C_\alpha=0.35$, $C_\Omega=33$,
$\kappa_{\rm f}=5$, $\epsilon_{\rm f}=1$, $\tilde{g}=3$.
}\label{Fpbutter}\end{figure}

As in the simulations of BBS, we have chosen negative values of
$\alpha_{\rm K}$, but this choice only affects the direction of propagation
of the dynamo waves. There are dynamo waves traveling in the positive
$z$-direction at $x=\pm\pi$ and in the negative $z$-direction at $x=0$,
which is consistent with the three-dimensional simulations. These waves are
best seen in a space-time (or butterfly) diagram; see \Fig{Fpbutter}. Note also
that there is an initial adjustment time during which the overall magnetic energy
settles onto its final value (consistent with resistively limited saturation)
and the cycle period increases by a small amount.

Compared with the one-dimensional model, the values of $Q^{-1}$ and
$\epsilon_{\rm m}$ are about 30\% larger in the two-dimensional model,
but $\omega_{\rm cyc}$ is about 3 times smaller. This may reflect the fact
that in the present geometry the upward and downward traveling
dynamo waves can propagate less freely,
because they are now also coupled in the $x$ direction.

For comparison, in the simulation of BBS,
the inferred input parameters for modeling purposes are
$k_{\rm m}^2=2$ (the field varies in $x$ and $z$),
$S/\eta k_1^2=2000$,
$R_{\rm m}\approx u_{\rm rms}/\eta k_{\rm f}\approx80$,
$C_\alpha\approx\tilde\kappa_{\rm f}=1...2$.
The resulting non-dimensional output quantities are
$B^2_{\rm fin}/B_{\rm eq}^2\approx30$, $Q^{-1}\approx0.02$,
$\epsilon_{\rm m}\approx0.11$, $\omega_{\rm cyc}/S=0.005\dots0.010$,
and $\lambda/(\eta k_{\rm m}^2)=30$.

\begin{figure}[t!]\centering\includegraphics[width=0.45\textwidth]{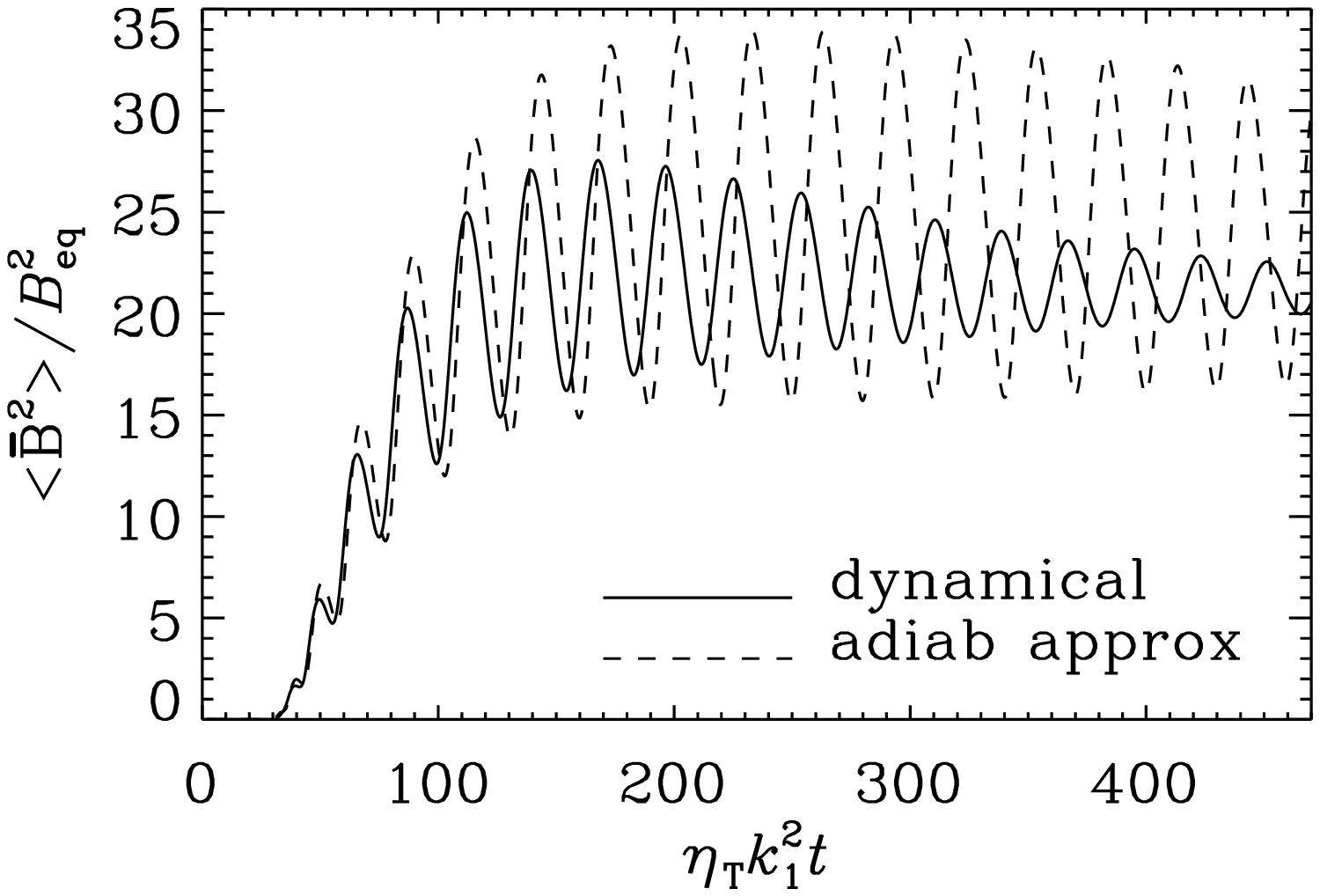}\caption{
%\begin{figure}[t!]\plotone{f5.eps}\caption{
Evolution of the large scale magnetic energy for Model~AG2
(solid line). The dotted line gives the comparison with the
corresponding adiabatic approximation (see appendix A).
$R_{\rm m}=100$, $C_\alpha=0.5$, $C_\Omega=20$,
$\kappa_{\rm f}=5$, $\epsilon_{\rm f}=1$, $\tilde{g}=3$.
}\label{Fpcomp}\end{figure}

Model~R1 gives, within a factor of two,
about the right saturation field strength, and also the
values of $Q^{-1}$, $\epsilon_{\rm m}$ and $\omega_{\rm cyc}/S$ agree
reasonably well with the simulations, but the kinematic growth rate is
too high. Also, the value of $R_{\rm m}$ is probably larger in
the simulation where we estimated $R_{\rm m}\approx80$.
In order to have the right growth rate, $C_\alpha$ has to
be lowered. In order to match then the right saturation field strength
we have to have $\tilde{g}\approx3$. One such case is Model~AG2,
where $R_{\rm m}=100$. Now cycle frequency, growth rate, as well
as $Q^{-1}$ agree reasonably well with the simulation.

The simulation of BDS is more resistive,
$S/\eta k_1^2=1000$ and $R_{\rm m}\approx30$,
but the resulting field strength is only somewhat smaller,
$B^2_{\rm fin}/B_{\rm eq}^2\approx20$, whereas $Q^{-1}\approx0.02$,
$\epsilon_{\rm m}\approx0.11$ and $\omega_{\rm cyc}/S=0.013\dots0.015$
are all somewhat enhanced relative to BBS.
Models s1\ldots s3 (where $\tilde{g}=1$) and S1\ldots S3
(where $\tilde{g}=3$) are now appropriate for comparison, because they
all have $S/\eta k_1^2=1000$. Model~S1 with $\tilde{g}=3$ gives the best
agreement for $B^2_{\rm fin}$, but the cycle frequency is too small.
For Model~s3 with $\tilde{g}=1$, $\omega_{\rm cyc}$ is about right,
but now $B^2_{\rm fin}$ is too small.

In all models the values of $b^2_{\rm fin}$ in table 5 are smaller than in
the simulations. As discussed in \Sec{Sfinal}, this is readily explained
by the fact that our model does not take into account small scale
dynamo action resulting from the non-helical component of the flow.

Comparing the two simulations with different values of $R_{\rm m}$
(BBS and BDS), the cycle frequency changes by a factor compatible
with the ratio of the two magnetic Reynolds numbers. This is not
well reproduced by a quenching expression for $\eta_{\rm t}$ that is
independent of $R_{\rm m}$ (case~I). On the other hand, if
$\eta_{\rm t}\propto\alpha$ (case~II), $\omega_{\rm cyc}$ becomes far
smaller than what is seen in the simulations. A possible remedy would
be to have some intermediate quenching expression for $\eta_{\rm t}$.
We should bear in mind, however, that our current model ignores
the feedback from the large scale motions. Such feedback is indeed
present in the simulations, which also show much more chaotic
behavior (e.g.\ Fig.~8 of BBS) than our model; see \Fig{Fpcomp}.
A more realistic model should therefore allow for more degrees
of freedom. In particular, the quenching should be allowed to
be nonuniform in space. This and other extensions of the model
are discussed in the next section.

\section{Possible extensions of the model}
\label{Sextent}

The dynamical quenching model allows us now to test a number of additional
aspects and properties that have been (or can be) seen in direct simulations.

\subsection{Cross helicity evolution and large scale velocity feedback}

Although we were justified in ignoring the small scale cross helicity
contribution to $\emf$, we found from the simulation of BBS that the
{\it large scale} cross helicity is non-negligible. This turned out to be
the result of the forcing function for the large scale velocity and the
asymmetry of the large scale field with respect to $x=0$, and thus with
respect to the large scale velocity. The correlation between the forcing
function for the large scale flow and the large scale magnetic field serves
as a driver in the cross helicity evolution equation.
In principle, we should explicitly
couple the equation for the cross helicity into the model. Our models with
imposed shear, however, did not produce the required symmetry breaking
that would lead to a significant contribution to the large scale cross
helicity. This is because we treat the large scale velocity as being
kinematic. This should be explored further in future work by accounting
for the dynamical feedback from the large scale motions.

\subsection{Antiquenching}

As long as the feedback from the small scale motions onto $\alpha$ and
$\eta_{\rm t}$ involve 
the quantity $\bra{\aaaa\cdot\bb}$, i.e.\ as long as \Eqs{dABdt}{dabdt}
remain fully coupled, \Eq{helicity_eqn} is guaranteed to be satisfied.
Thus, the magnetic helicity equation is obeyed regardless of how
$\bra{\aaaa\cdot\bb}$ is coupled to $\alpha$ or $\eta_{\rm t}$.
It may be that under certain conditions, $\alpha$ and $\eta_{\rm t}$
may even increase with increasing field strength, which we refer to as
`anti-quenching'. Brandenburg, Saar, \& Turpin (1998) used such models
to explain the increase of relative stellar cycle frequency with
increasing field strength. As an illustrative example, we have
considered the case $\alpha=\alpha_{\rm K}-\alpha_{\rm M}$, i.e.\ with
the opposite sign as in \Eq{alpha}, and
$\eta_{\rm t}=\eta_{\rm t0}(1+\bra{\bb^2}^2/B_{\rm eq}^2)$,
where $\bra{\bb^2}$ is linked to $\bra{\aaaa\cdot\bb}$ via
\Eqs{kftilde}{kf2}.
Note that $\eta$ antiquenching goes with the 4th power, so
that it eventually dominates over $\alpha$ antiquenching.
The resulting evolution of $\bra{\meanBB^2}$
shows the expected resistively limited saturation phase.
Alternatively, if $\alpha$ is made to increase with increasing
small scale field strength, the resulting large scale field
can saturate faster than usual. A similar behavior can be modeled
by choosing $\tilde{g}<0$, which also speeds up the initial build-up
of large scale magnetic energy. This is possible, and consistent with
the magnetic helicity equation, because the initial build-up of
$\bra{\meanBB^2}$ happens in this case simultaneously with a sharp
burst in $\bra{\bb^2}$ such that the sum of $\bra{\meanAA\cdot\meanBB}$
and $\bra{\aaaa\cdot\bb}$ is still approximately constant.

\subsection{Magnetic buoyancy}

In solar and galactic 
dynamo theory the possibility of rising magnetic flux tubes
contributing to the $\alpha$ effect has been discussed (Leighton 1969;
Ferriz-Mas, Schmitt, \& Sch\"ussler 1994; Hanasz \& Lesch 1997;
Brandenburg \& Schmitt 1998; Moss, Shukurov, \& Sokoloff 1999; Thelen 2000; 
Spruit 2002). For the sun,  the idea
is that flux tubes emerge from the toroidal magnetic field belt at the
bottom of the convection zone and become twisted by the Coriolis force.
We point out that \Eq{fullset2} does already capture part of this effect.
If there is a strong partly buoyant magnetic field at the bottom of the
convection zone, it would contribute to $\bra{\meanJJ\cdot\meanBB}$
and therefore, through \Eq{fullset2}, to $\alpha$. Of course, this
effect cannot constitute a dynamo on its own as there is no source of
magnetic energy. However, in conjunction
with shear, from which energy can be tapped, this effect
could lead to dynamo action. Modeling this in the framework of dynamical
quenching would be a suitable way
to include the effects of magnetic buoyancy such
that magnetic helicity conservation is obeyed.

\subsection{Oscillatory imposed fields}

If there is a uniformly imposed magnetic field that is oscillatory
in time, \Eq{fullset2} would predict that $\alpha$ is also oscillatory.
If the oscillation frequency is high enough, the adiabatic approximation
breaks down. One might wonder whether this would be a way to test
explicitly the dynamical $\alpha$ quenching concept numerically.
However, it turns out that the resulting averaged $\alpha$ is even
smaller on average than what is predicted based on the
adiabatic approximation.
Thus, although dynamical quenching enhances dynamo generation in the
case of self-generated fields, it actually lowers $\alpha$ in the
presence of imposed oscillatory fields.

\subsection{Selective decay}

In is interesting to note that \Eq{fullset2} can also be applied to
the case of decaying magnetic fields. In that case, it predicts
reduced turbulent decay if $\bra{\meanJJ\cdot\meanBB}\neq0$. Thus,
accordingly, a fully helical magnetic field should only decay at
the resistive rate, whereas a non-helical magnetic field would decay
at the turbulent diffusive rate. The slow decay of helical fields
is well known and leads to the so-called Taylor states where
magnetic helicity is maximized and magnetic energy minimized
(e.g.\ Montgomery, Turner, \& Vahala 1978).

\subsection{Hyperdiffusion}

Numerical experiments allow one to understand the physics described by
the equations by modifying certain terms. Particularly enlightening has been
the use of hyperresistivity (or hyperdiffusion) by which the ordinary diffusion operator,
$\eta\nabla^2$, is simply replaced by $\eta_2\nabla^4$.
The diffusion at small scales is usually fixed by the mesh resolution and
kept unchanged, but with hyperdiffusion the diffusion at large scales can
be decreased substantially. This method is
frequently used in turbulence research, but the effects on helical
dynamos are quite striking: the saturation field strength is considerably
enhanced and the saturation phase prolonged (Brandenburg \& Sarson 2002).
Our model reproduces these features 
if $\eta k_{\rm f}^2$ is replaced
by $\eta_2 k_{\rm f}^4$, $\eta\meanJJ$ is replaced by $-\eta_2\nabla^2\meanJJ$,
and $R_{\rm m}=\eta_{\rm t}/(\eta_2 k_{\rm f}^3)$ is used. The simulations
of Brandenburg \& Sarson (2002) showed (for the helical dynamo without shear)
that the large scale field behavior
depends on the diffusion at the scale of the large scale field itself and
not, as one might naively expect, on the diffusion at small scales.
This behavior is clearly reproduced by the dynamical quenching model:
reducing the {\it microscopic} $\eta$ in the mean field equation (and not
in the dynamical quenching equation) increases saturation time and saturation
value as expected. This can be taken as additional validation of the
dynamical quenching model.

\subsection{Losses of small scale field}

Open boundaries may provide a means of shedding magnetic helicity
and thereby alleviating the magnetic helicity constraint
(Blackman \& Field 2000; Kleeorin et al.\ 2000; 2002). Numerical simulations
have shown, however, that when no additional boundary physics
is imposed to transport preferentially  quantities of a particular scale, 
most of the magnetic helicity is lost by the
large scale field (Brandenburg \& Dobler 2001). In this case, 
the growth of the large scale field cannot be accelerated. 
In order to check whether accelerated growth of
large scale fields is at least in principle possible we have modeled the
preferential 
shedding of small scale fields in two different ways, both with similar
results. Adding an overall loss term of the form $-\alpha_{\rm M}/\tau_{\rm loss}$
on the right hand side of \Eq{fullset2} leads to substantial increase of
the large scale field [note that this is distinct from the
$-\alpha_{\rm M}/T$ in Kleeorin et al.\ (2000), which is just the same as
our second term in \Eq{dynquench}].
Likewise, setting $\bra{\bb^2}$ (and hence
$\alpha_{\rm M}$) to zero in sporadic
intervals accelerates the growth phase and enhances the saturation value.
Similar results have meanwhile also been obtained by restarting Run~3 of
B01 after sporadically removing magnetic field at and below the forcing scale
(see Fig.~14 of BDS).
This confirms an important prediction from the dynamical quenching model.

\subsection{Generalization to nonuniform $\alpha$}

Our approach is based on magnetic helicity which is a volume integral.
However, in astrophysical bodies kinetic and magnetic helicities
are not spatially constant and change sign at the equator.
Generalizing \Eq{dynquench} to the case of space dependent
$\alpha_{\rm K}$ and $\alpha_{\rm M}$
seems at first glance straightforward: omit the angular brackets and replace
$\dd/\dd t$ by $\partial/\partial t$, as was done already in the early
work of Kleeorin \& Ruzmaikin (1982).
It may also be necessary to include a local
phenomenological magnetic helicity flux
transport term, for example of the form $\eta_\alpha\nabla^2\alpha_{\rm M}$
(in addition to whatever global flux terms may be present,
e.g. Blackman \& Field 2000; Kleeorin et al.\ 2002).
In the presence of large scale
(meridional) flows, it may furthermore be appropriate to use the advective derivative,
$\DD/\DD t=\partial/\partial t+\meanUU\cdot\nab$.
However, the magnetic helicity density, $\aaaa\cdot\bb$, is not gauge-invariant
and it is no longer strictly related to $\jj\cdot\bb$ locally. The hope would
be that the generalization outlined above may still be useful as an
approximation.

We have performed calculations with
$\alpha_{\rm K}=\alpha_{\rm K0}\sin k_1z$ and
$C_\alpha\equiv\alpha_{\rm K0}/(\eta_{\rm T}k_1)=5$.
The resulting large scale field strength is approximately
equal to $B_{\rm eq}$, and depends only weakly on $R_{\rm m}$.
This is consistent with Kleeorin et al.\ (2002).
Simulations with the same sinusoidal $\alpha$ profile
(Brandenburg 2001b) have shown that the resulting
large scale field varies mostly in the $x$ direction,
which is incompatible with the present model. Also the resulting
field strength was actually significantly below $B_{\rm eq}$.

In the presence of open boundaries the present model predicts large scale
field strengths that decrease inversely proportional with $R_{\rm m}$.
This is even steeper that what was found in the simulations
(Brandenburg \& Dobler 2001). Thus, in its present form the dynamical
quenching model does not reproduce satisfactorily the numerical results
when $\alpha$ varies in space. However, with the help of simulations
it should be possible to identify which of the steps in the derivation
of dynamical $\alpha$ quenching are no longer satisfied, and hence
what the cause of the problem is.

\section{Conclusions}
\label{Sconcl}

The magnetic helicity evolution equation is a constraint
that must be satisfied by any dynamo theory. 
When we apply this in the mean field formalism
with the prescription that the $\alpha$ effect 
is proportional to the difference between kinetic and current helicities,
dynamical $\alpha$ quenching emerges as the only theoretically consistent 
approach to $\alpha$ quenching.  This is supported by
comparisons with numerical simulations of dynamos with and without shear. 
Fixed-form, algebraic quenching prescriptions may apply in a specific parameter
regime (e.g. the saturated phase),
 but are invalid for earlier times and are inconsistent
with results from  time dependent analyses. Only dynamical
quenching has predictive power.

A key result from dynamical quenching is that 
near-equipartition large scale field strengths are reached independently
of the magnetic Reynolds number by the end of the kinematic phase.
Final saturation is only reached on a resistively limited
($R_{\rm m}$ dependent) time scale but with 
a saturation value independent of $R_{\rm m}$ and equal
to $B_{\rm
fin}^2/B_{\rm eq}^2=\tilde{k}_{\rm f}/\tilde{k}_{\rm m}$; see
\Eqs{final_field_strength}{final_state}.

Although magnetic helicity conservation provides
a basis for a dynamical quenching of $\alpha$, the form of $\eta_{\rm t}$
must be prescribed at present.  We have shown that current simulations
of $\alpha^2$ (shear-free) dynamos constrain the dynamical quenching of 
$\meanemf\cdot{\overline {\bf B}}$ which is a combination of
$\alpha$ and  $\eta_{\rm t}$, but they do not separately constrain
$\eta_{\rm t}$. On the other hand, cycle
periods, emerging only in dynamos with shear, can.  
At present, the shear dynamo simulations are best
described by a dynamical quenching theory in which $\eta_{\rm t}$ is
only weakly dependent on the magnetic field; $\tilde{g}\approx3$ in \Eq{g}.
Higher resolution simulations are needed to verify this.

The two-scale dynamical nonlinear quenching approach
based on magnetic helicity conservation
discussed herein constitutes an improvement over
fixed-form algebraic quenching approaches.
Nevertheless, there are aspects of high-Reynolds number
$\alpha\Omega$ dynamos that may require the  
theory to be augmented. 
For example, we have only considered a spatially uniform
$\alpha$ coefficient. Allowing for spatial gradients
in $\alpha$ will introduce local helicity flux terms
that are important for astrophysical bodies where $\alpha$ changes
sign across the equator.  In addition, helicity flux
across global boundaries was also ignored in our calculation, 
though we know that real systems have boundaries.
The associated  boundary magnetic 
helicity flow may be important in coupling the dynamo growth to
magnetic helicity evolution.

\acknowledgments
We thank D.\ Moss and D.\ Sokoloff for constructive comments and suggestions.
We acknowledge the hospitality of the Aspen Center for Physics
and the Institute for Theoretical Physics
at the University of California, Santa Barbara, where much of this work
was carried out.
This research was supported in part by the National Science Foundation
under Grant No. PHY99-07949. EB also acknowledges support from
DOE grant DE-FG02-00ER54600.

\appendix
\section{The adiabatic approximation}

We discuss here the justification for when the time derivative
in the dynamical quenching expression can be neglected
(the adiabatic approximation).

In the steady state, $\alpha_{\rm M}$ and $\bra{\meanBB^2}$
are given by \Eqs{alpM_final}{final_field_strength}, respectively.
Linearizing \Eqs{fullset1}{fullset2} about this state yields
\EQ
\half{\dd\qq\over\dd t}=
\pmatrix{
-(\alpha_{\rm K}/\tilde{k}_{\rm m}-\eta_{\rm t})k_{\rm f}^2 & \eta k_{\rm f}^2\cr
-(\alpha_{\rm K}/\tilde{k}_{\rm m}-\eta_{\rm T}\!)k_{\rm m}^2 & 0 }\qq,
\label{adjust}
\EN
where $\qq\equiv\left(\delta\!\ln\alpha_{\rm M},\,
\delta\!\ln\bra{\meanBB^2}\right)$ is the state vector
for the logarithmic departure from equilibrium.
For excited solutions, the terms in the first column
of the matrix in \eq{adjust} are usually positive,
even for $\alpha\Omega$ dynamos.
Inspecting the diagonal terms
shows that near the saturated state,
$\alpha_{\rm M}$ is adjusting rapidly
on a dynamical time scale whilst $\bra{\meanBB^2}$
is marginal and adjusts only indirectly (via $\alpha_{\rm M}$)
on a resistive time scale. We can therefore use
the adiabatic elimination principle (e.g.\ Haken 1983) to
remove the explicit time dependence of $\alpha_{\rm M}$
by replacing \Eq{fullset2} by
\EQ
0=\left(\alpha\bra{\meanBB^2}
-\eta_{\rm t}\mu_0\bra{\meanJJ\cdot\meanBB}\right)/B_{\rm eq}^2
+\alpha_{\rm M}/R_{\rm m}.
\label{fullset2_stat}
\EN
Substituting $\alpha_{\rm M}=\alpha-\alpha_{\rm K}$ and solving for $\alpha$
leads to \Eq{Gruz+Diam}.

The adiabatic approximation corresponds to the limit in which
memory effects become negligible. This is best seen by considering
the integral form of \Eq{fullset2},
\EQ
\alpha=2\eta k_{\rm f}^2\int_0^{\displaystyle{t}}
G(t,t')\left(\alpha_{\rm K}+R_{\rm m}\eta_{\rm t}\mu_0
{\bra{\meanJJ\cdot\meanBB}\over B_{\rm eq}^2}\right)\dd t',
\label{greens1}
\EN
with the Green's function
\EQ
G(t,t')=\exp\left[-2\eta k_{\rm f}^2\int_{\displaystyle{t'}}^{\displaystyle{t}}
\left(1+R_{\rm m}{\bra{\meanBB^2}\over B_{\rm eq}^2}\right)\dd t''\right].
\label{greens2}
\EN
As long as the field is weak, the width of the Green's function is
the resistive time scale, but when $\bra{\meanBB^2}/B_{\rm eq}^2$
is of order unity the large $R_{\rm m}$ factor becomes important
and the width of the Green's function reduces to a dynamical
time scale. In that case, the $\meanBB$ dependent terms in brackets
can be pulled out of the integrals in \Eqs{greens1}{greens2},
in which case \Eq{Gruz+Diam} is recovered.

\end{document}